\documentclass{article}
\usepackage[utf8]{inputenc}
\usepackage{graphicx}
\usepackage{float}
\usepackage{authblk}
\usepackage{amsfonts}
\usepackage{setspace}
\usepackage{url}
\usepackage{stfloats}
\usepackage{multicol}
\graphicspath{ {./Figures/} }

\usepackage[numbers,sort&compress]{natbib}

\usepackage[usenames, dvipsnames]{xcolor} 
\usepackage{changes}
\definechangesauthor[color=red]{commentvld}
\definechangesauthor[color=lightgray]{style}
\setauthormarkuptext{}

\usepackage{todonotes}

\usepackage[margin=0.75in]{geometry}

\usepackage{booktabs}
\usepackage{adjustbox}

\date{\today}

\title{An Accurate and Transferable Machine Learning Potential for Carbon}

\author[1]{Patrick Rowe}

\author[2]{Volker L. Deringer}

\author[1]{Piero Gasparotto}

\author[3]{G\'abor Cs\'anyi}

\author[1,a]{Angelos Michaelides}

\affil[1]{Thomas Young Centre, London Centre for Nanotechnology, and
Department of Physics and Astronomy, University College London,
Gower Street, London, WC1E 6BT, U.K.}

\affil[2]{Department of Chemistry, Inorganic Chemistry Laboratory, University of Oxford, Oxford OX1 3QR, U.K.}

\affil[3]{Engineering Laboratory, University of Cambridge, Trumpington Street, Cambridge CB2 1PZ, U.K.}

\affil[a]{Author to whom correspondence should be addressed: angelos.michaelides@ucl.ac.uk}

\begin{document}\sloppy

\maketitle

\begin{abstract}
    We present an accurate machine learning (ML) model for atomistic simulations of carbon, constructed using the Gaussian approximation potential (GAP) methodology. 
    The potential, named GAP-20, describes the properties of the bulk crystalline and amorphous phases, crystal surfaces and defect structures with an accuracy approaching that of direct \textit{ab initio} simulation, but at a significantly reduced cost.
    We combine structural databases for amorphous carbon and graphene, which we extend substantially by adding suitable configurations, for example, for defects in graphene and other nanostructures. 
    The final potential is fitted to reference data computed using the optB88-vdW density functional theory (DFT) functional. 
    Dispersion interactions, which are crucial to describe multilayer carbonaceous materials, are therefore implicitly included.
    We additionally account for long-range dispersion interactions using a semianalytical two-body term and show that an improved model can be obtained through an optimisation of the many-body smooth overlap of atomic positions (SOAP) descriptor. 
    We rigorously test the potential on lattice parameters, bond lengths, formation energies and phonon dispersions of numerous carbon allotropes. 
    We compare the formation energies of an extensive set of defect structures, surfaces and surface reconstructions to DFT reference calculations. 
    The present work demonstrates the ability to combine, in the same ML model, the previously attained flexibility required for amorphous carbon [\textit{Phys. Rev. B}, \textbf{95}, 094203, (2017)] with the high numerical accuracy necessary for crystalline graphene [\textit{Phys. Rev. B}, \textbf{97}, 054303, (2018)], thereby providing an interatomic potential that will be applicable to a wide range of applications concerning diverse forms of bulk and nanostructured carbon.
\end{abstract}

\vspace{1cm}
\twocolumn


\section{Introduction}
The same characteristics which make carbon a fascinating element for study also make it challenging to model computationally. 
It exhibits some of the greatest structural diversity - and associated diversity of properties - of any of the elements \cite{Novoselov2004,Kroto1985,Martinez-Canales2012,Powles2009,Deringer2017a,Tomnek2014,Mirzayev2017}.
Its allotropes range from zero to three-dimensional, have metallic, semiconducting and insulating phases and boast mechanical properties including some of the highest tensile strengths, hardnesses and bulk moduli measured \cite{Lee2008,Neves2001}. 
It is unsurprising, therefore, that carbon is considered to be not just an element of prime technological importance, but also remains the subject of continued fundamental scientific study \cite{Geim,Avouris2007,Yan2015,Zhang2016,Zhang2019,Yang2015a,Yuan2016}.

Current applications of elemental carbon are numerous, they include lightweight and strong structural materials, anodes of batteries and components in advanced optical technologies \cite{Neves2001,Yuan2016,Zhang2016,Arico2005,Kholmanov2015,Bonaccorso2010}. 
The usefulness of elemental carbon has also clearly not yet been exhausted; future applications propose to make use of graphene’s unique electronic properties for advanced electronics \cite{Avouris2007,Neto2009}, carbon nanotubes’ structural and optical characteristics for performance materials \cite{Kholmanov2015,Farrera2017} and the thermal and optical properties of diamonds for laser optics \cite{Neves2001,Courvoisier2016,Latawiec2015}, amongst innumerable other examples.
Atomistic simulations have played a major role in developing our understanding of carbon materials.
Among the many scientific problems that have been addressed with carbon potentials, we may mention the wear process of diamond \cite{Pastewka2011} or the compression behaviour of glassy carbon \cite{Shiell2018}.
The first many-body interatomic potential for modelling carbon was published in 1988 by Tersoff.
This potential was used to investigate the properties of carbon's crystalline and amorphous allotropes \cite{Tersoff1988}.
The reactive empirical bond order (REBO, REBO-II) potentials built on the original Tersoff formulation to include a wider range of parameters and data in the fit, as well as adding additional conjugation and torsional terms, and modifying the bond order expression for small angles \cite{Brenner1990,Connor2015}.
However, the Tersoff and REBO-II potentials only considered nearest neighbour interactions and did not account for the effects of dispersion.
The adaptive intermolecular REBO (AIREBO) potential \cite{Stuart2000} sought to correct this by adding an additional long-range Lennard-Jones term between atoms with larger separations, while making no modifications to the short-range part of the potential.
The long-range carbon bond order potential (LCBOP) not only increased the range of the potential to account for longer-range interactions but also constituted a complete reparameterisation of the bond order potential to improve the accuracy and transferability of the model, though long-range dispersion interactions are still omitted \cite{Los2003}. 
Further developments, beyond the scope of detailed discussion here, include the LCBOP-II potential which expanded the application range of the model to include the liquid phase \cite{Ghiringhelli2005,Los2006}, the environment dependant interatomic potential (EDIP) for carbon \cite{Marks2006} which employed properties calculated from \textit{ab initio} simulation in its parameterisation, the introduction of a dynamic cut-off to bond-order potentials \cite{Pastewka2008} and a recent reparameterisation of a carbon ReaxFF potential \cite{Srinivasan2015}. 
Notwithstanding the long-standing success of these potential models, there are inherent limitations to even the most advanced of them.
Such issues are particularly relevant when one departs from the idealised structures (diamond, graphite, {\em etc.}), as shown in two detailed benchmark studies by de Tomas et al. \cite{DeTomas2019,Tomas2016}.
%

%
%
%
%
%
Machine learning (ML) has recently arisen as a way of addressing some of these limitations.
A number of practical approaches for modelling the potential energy surface (PES) using ML have been developed in recent years, employing algorithms including artificial neural networks, Gaussian process regression, and compressed sensing \cite{Parrinello2007,Bartok2010,Schutt2018,Huan2017,Bartok2017,Kolb2017,Dolgirev2016,Glielmo2017}.
The demonstrated ability of ML algorithms to fit arbitrary functions with an extremely high accuracy \cite{Korkov1992}, combined with recent developments in high-dimensional descriptors for atomic systems makes ML approaches for the development of interatomic potentials an increasingly popular approach \cite{Bartok2017,Handley2014,Behler2011,Behler2016,Deringer2019}. 
Indeed, the use of machine learning methodologies to model carbon has a significant precedent.
Some of the first examples of ML potentials, using both Gaussian process regression and artificial neural networks, were fitted for graphite and diamond, these were tested with regards to properties off the crystalline phases and the graphite-diamond phase coexistence line \cite{Parrinello2007,Bartok2010,Khaliullin2010}. 
One of the first neural-network potentials was used for large-scale simulations of the diamond nucleation mechanism \cite{Khaliullin2011}.
We recently introduced a machine learning potential for pristine graphene constructed using the Gaussian approximation potential (GAP) framework, which achieved excellent accuracy when benchmarked against DFT and experiment for a wide range of lattice and dynamical properties, including the phonon dispersion relations, thermal expansion and Raman spectra at different temperatures \cite{Rowe2018}.
While achieving good accuracy in a specific region of configuration space is not trivial, the problem of the transferability of a potential is much more challenging to solve from a ML perspective.
In 2017, some of us reported a highly transferable GAP model trained primarily on the amorphous and liquid phases of carbon (henceforth termed GAP-17), based on DFT-LDA reference data. The focus, there, was somewhat complementary---to be able to describe very diverse structural environments, albeit accepting a degree of numerical error. As an example, the in-plane force errors for a pristine graphene sheet are 0.03 eV \AA{}$^{-1}$ with the graphene-only GAP mentioned above, as compared to 0.27 eV \AA{}$^{-1}$ with GAP-17. For comparison, these errors for a range of commonly used empirically fitted potentials range from 0.6 to 3.1 eV \AA{}$^{-1}$ (more details are in Ref. \cite{Rowe2018}).
In return, owing to the flexibility and transferability ensuing from its choice of reference database, GAP-17 enabled the study of a number of scientific problems which involve diverse structural environments, including understanding the mechanism of growth of $\mathrm{sp}^3$ hybridised amorphous carbon by ion deposition \cite{Caro2018}, extensive studies of the surface properties (and chemical reactivity) of tetrahedral amorphous carbon \cite{Deringer2018b,Caro2018,Caro2018a}, the structure of ``porous'' carbonaceous materials which are relevant to applications in batteries and supercapacitors \cite{Deringer2018a,Huang2019}, and crystal-structure prediction \cite{Deringer2017a}.

\begin{figure*}[tp]
    \centering
    \includegraphics[width=15cm]{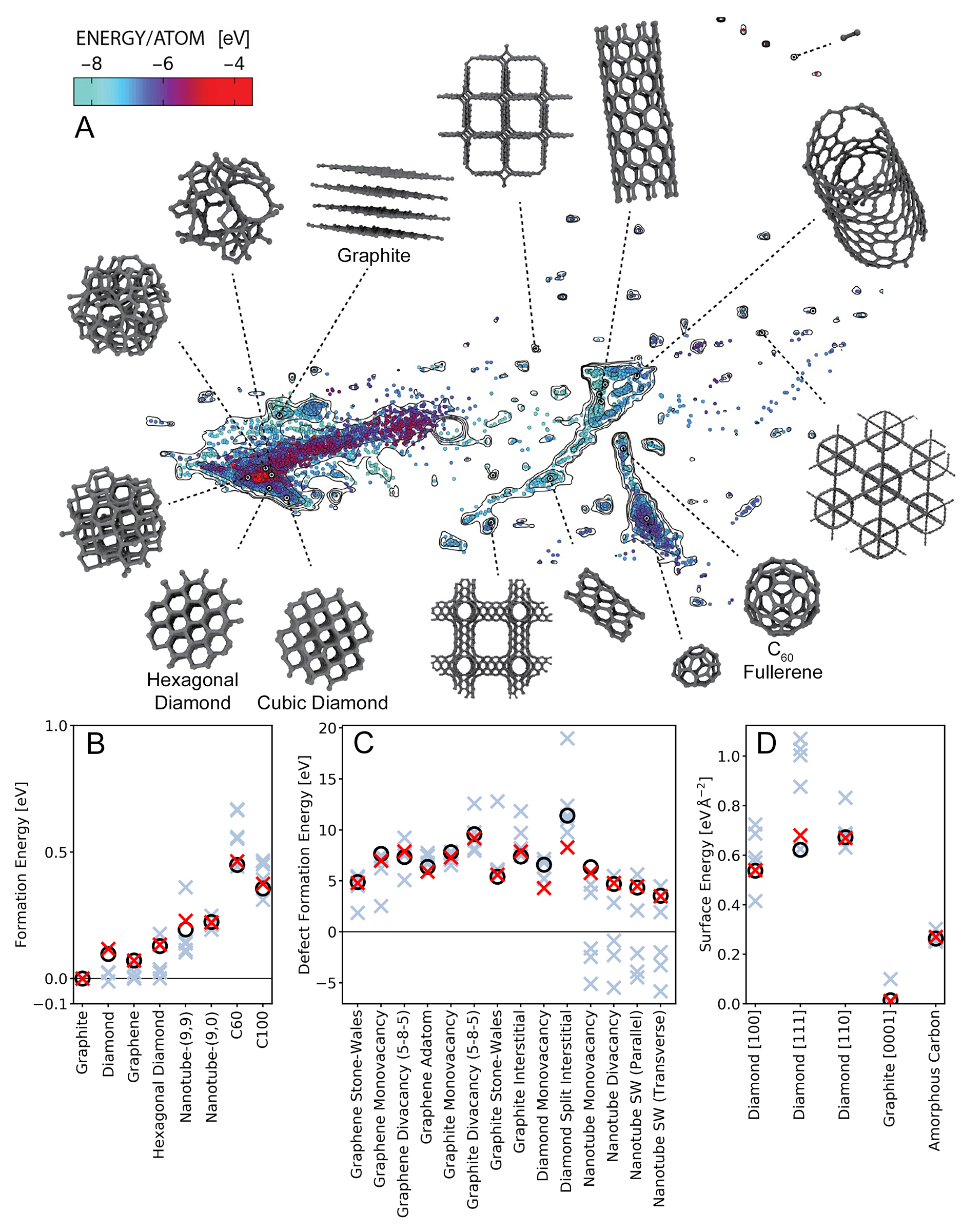}
    \caption{Overview of some of the key structures included in the training shown through a sketch-map representation (top) as well as selected information on the performance of the potential for a variety of properties. (a) Sketch-map representation of the total data set for carbon generated as part of this work. Select structures are identified for graphite, diamond, hexagonal diamond (Lonsdaleite), amorphous carbon and fullerenes. Points are coloured according to their energy, while contours indicate the density of the database population in a particular region. Bottom, is a summary of (b) the predicted crystalline formation energies, (c) defect formation energies and (d) surface energies, comparing the DFT (optB88-vdW) reference (black circles)  GAP-20 (red crosses) and all other models (blue crosses).}
    \label{fig:sketchmap}
\end{figure*}

The model we present here, GAP-20, builds on all of the previous work applying the GAP machine learning methodology to the development of carbon potentials, to achieve the accuracy required for capturing subtle differences in formation energies of nanostructures or in defect formation energies, and for describing phonon dispersions to within meV accuracy - while maintaining the flexibility and transferability of GAP-17. 
Importantly, all data are generated using a dispersion-corrected DFT method which properly accounts for longer-range interactions in low-dimensional carbon structures, and the fitting architecture is adapted to account for those.
Our tests suggest GAP-20 to be suitable as a ``general-purpose'' carbon ML potential for diverse areas of study.
While detailed discussions of the construction and testing of the potential will be given in subsequent sections, we take a moment to highlight the main points here.
The composition of the training data set and performance of this potential is summarised in figure \ref{fig:sketchmap}.
GAP-20 correctly predicts the formation energies of diamond, graphite, fullerenes and nanotubes, to an accuracy of a few meV, and achieves comparable accuracy for a number of crystalline and amorphous surfaces.
The computed formation energies of defects are also accurate, with overall errors significantly lower than those obtained from comparable empirical models.
At the same time, GAP-20 can accurately predict the behaviour of high temperature liquid carbon over a wide range of temperatures and densities, which will be shown below.
We believe that these features make GAP-20 a useful tool for the accurate modelling of nanostructured carbons; nanotubes, graphitised carbon and materials with varying degrees of defects and disorder.
The rest of the paper will be organised as follows. 
We first describe our process for the construction of a training set suitable for developing such a potential.
We then give details on the construction and training of the model itself, with discussion of particular aspects which required special attention or optimisation.
Subsequently, we present an extensive and rigorous testing of our model, for a wide range of properties. 
We also compare the results of our model to a selection of commonly used empirical potentials which model the interatomic interactions in carbon with differing degrees of simplification.
Specifically we choose the Tersoff, REBO-II, AIREBO and LCBOP models. 
%
%
The selection of potentials considered here is by no means exhaustive and is only intended to give some basis for comparison between previous work and the model we introduce, as well as illustrating how the inclusion or exclusion of different interactions (e.g. dispersion interactions) may affect the performance of a model.
A more detailed benchmarking across a wider range of potentials, complementing the existing detailed tests for amorphous and ``graphitised" carbons \cite{DeTomas2019}, may be the subject of future work.
%

%
%

%
%
\section{Generation and Selection of Training Data}
One of the challenges inherent in constructing a generalised potential for carbon is the enormous variety of structures which must be considered.
In addition to its more commonly encountered crystalline phases; diamond and graphite, carbon may be found in forms of differing dimensionality, from zero dimensional fullerenes, to one dimensional nanotubes, two dimensional graphene and three dimensional amorphous forms \cite{Tomnek2014}.
%
%
%

%
Specifically in the case of ML, one is drawn to the problem of the composition of the large database of example configurations, known as the training data set. 
For a potential to be both accurate and transferable, its training data set ought to include representative configurations from all of the thermally accessible chemical space.
One might initially suggest that the problem is therefore intractable, if in order to produce a potential which is capable of accurately modelling all of the relevant phases of carbon, we must explore the entirety of the vast 3N-dimensional chemical space. 
It is an empirical observation, however, that the thermally accessible and physically relevant regions of this chemical space constitute a vastly reduced subset of all of the available configurations \cite{Bartok2018,Pickard2011a,Wales2004}.
Further, rather than an exploration of the 3N dimensional space, in fitting the parameters of a ML algorithm we are primarily concerned with an exploration of the reduced dimensionality descriptor space \cite{Deringer2017,Bartok2018,Bartok2013a,Behler2017}.
In the case of atom centred descriptors such as the smooth overlap of atomic positions (SOAP), this represents the local environment around a particular atom rather than the global structure \cite{Bartok2013a}.
While the structural variability of carbon is globally almost infinite, many of these structures are constructed from similar local motifs, for example the tetrahedral building blocks of diamond \cite{Gasparotto2018,Wales1987}.
%
%
%
Similar logic may be applied to more complex structures.
The reference configurations which comprise our structural database are drawn from a wide variety of sources.
Regardless of the origin of the configuration itself (e.g., from the GAP-17 database), its properties, those being the total energy, atomic forces and virial stresses, which comprise the actual training data, are always computed using the same level of tightly converged plane-wave DFT including dispersion corrections. 
We use the VASP plane-wave DFT code, we perform spin-polarised calculations with the optB88-vdW dispersion inclusive exchange-correlation functional \cite{Lee2010,Bowler2011,Langreth2004,Klimes2010}, a plane-wave cut-off of 600 eV and a projector augmented wave pseudopotential \cite{Kresse1996,Kresse1996a,Kresse1999}.
A Gaussian smearing of 0.1 eV is applied to the energy levels and dense reciprocal space Monkhorst-Pack grids are used \cite{Monkhorst1976}. 
In the case of the reduced dimensionality allotropes; graphene and nanotube structures, the reciprocal space sampling is only performed in the directions in which the allotrope is periodic.
The properties of fullerene structures are calculated at the gamma point. 
For this potential, we choose the optB88-vdW functional as it has already been demonstrated to provide an excellent description of carbonaceous materials, in particular graphitic carbon – for which its prediction of the binding energy and interlayer spacing is in good agreement with experimental values \cite{Graziano2012}.
%
%

%
The database of configurations presented here uses as its foundation a combination of the training data sets for the two carbon potentials previously published primarily for liquid and amorphous carbon (GAP-17), and for pristine graphene, respectively \cite{Deringer2017,Rowe2018}.
%
%
%
%
A large number of new configurations are considered in addition to these existing ML data sets \cite{Hoffmann2016,Deringer2017a,Tomnek2014}.
We endeavour to comprehensively cover all the possible crystalline phases of carbon found at moderate temperatures and pressures, including more exotic allotropes.
To that end, DFT optimised structures for graphite, graphene, cubic and hexagonal diamond are included, as well as the structures of a library of fullerenes comprised of fewer than 240 atoms and all nanotube structures with chiral indices, $\mathrm{3 \leq n}$, $\mathrm{m \leq 10}$ with fewer than 240 atoms in their unit cell.
Optimised structures are also included for the SACADA database of exotic carbon allotropes \cite{Hoffmann2016} and the results of a GAP-17 driven random structure search \cite{Deringer2017a}.
In addition to bulk or pristine phases, the structures of relevant low Miller-index faces of the crystalline phases are included, along with a large number of important defect structures \cite{Li2005,Xu2018,Charlier2002,Skowron2015,Ristein2006,Kern1998,Ooi2006,Kern1998a}.
For all of these structures, we have performed some \textit{ab initio} and some iteratively improved GAP driven molecular dynamics simulations at a number of temperatures so as to also sample the region of phase space close to these local minima \cite{Rowe2018,Deringer2017}. 
The resulting database is comprised of ca. 17000 configurations each containing from 1 to 240 atoms per cell. 
The choice of which structures might be important for training a potential requires for the most part chemical or physical intuition on the part of the researcher \cite{Deringer2018,Bartok2017,Bartok2018}.
Some of these choices may be clear, for example the need to include configurations representing the bulk structures of diamond and graphite.
Others, however, such as the inclusion or exclusion of particular defect or surface structures, will depend on the desired application of the potential (and, to some extent, on personal choice).
To maximise the transferability of our model, we have produced as comprehensive a database as possible – too large to train on with current computational facilities. 
Rather than using the full database for sparsification, as commonly done in GAP fitting (including in the development of GAP-17), we instead allow the bulk of our training configurations to be chosen from the total dataset using a sampling method known as farthest point sampling (FPS) \cite{De2016,Bartok2017}. 
Within this set, we then carefully check the data saturation of our training with respect to the number of sparse points, which is discussed in section 3. 
%

%
This method allows us to start with a much more comprehensive database than previously, while still keeping the computational effort at the fitting stage tractable.
We wish for our training dataset to have the widest possible sampling of descriptors and forces – leaving no physically relevant configurations unsampled, while avoiding over-representation of particular regions of phase space.
%
%
FPS facilitates this, by allowing a selection of frames to be made based on a measure of the global similarity (in descriptor space) between possible configurations \cite{De2016,Bartok2017}.
Given a set of $\mathrm{n}$ descriptors of type $\mathrm{d}$ for a number of frames, $ \mathbb{Q} = \{\mathrm{\mathbf{q}}^{\mathrm{d, avg}}_{i=1 \dots \mathrm{n}}\}$, which are themselves the average of the individual descriptors of the atoms in a particular frame $\mathrm{\mathbf{q}}^{\mathrm{d}}_{\mathrm{i}}$, the FPS algorithm selects configurations so that at each step, the kernel distance between previously selected configurations $\mathbb{Q}_{\mathrm{selected}} = {\{\mathrm{\mathbf{q}}^{\mathrm{d, avg}}_{1} \ldots \mathrm{\mathbf{q}}^{\mathrm{d, avg}}_{\mathrm{m}}}\}$ and the new configuration $\mathrm{\mathbf{q}}^{\mathrm{d, avg}}_{\mathrm{m+1}}$ is maximised. That is,

\begin{equation}
    \mathrm{\mathbf{q}}^{\mathrm{d, avg}}_{\mathrm{m+1}} = \mathrm{argmax}_{\mathrm{\mathbf{q}}^{\mathrm{d, avg}}}[\mathrm{D}(\mathbb{Q}_{\mathrm{selected}}, \mathrm{\mathbf{q}}^{\mathrm{d, avg}})],
\end{equation}

\noindent
where $\mathrm{D}$ is the kernel distance between the selected descriptors (and associated frames) $\mathbb{Q}_{\mathrm{selected}}$ and the candiate descriptor $\mathrm{\mathbf{q}}^{\mathrm{d, avg}}$.
In our case, we use the SOAP descriptor as a structural fingerprint of a configuration and the dot product between two SOAP descriptors as our kernel similarity measure \cite{Bartok2013a}. 
As has previously been shown for molecular systems, we find that this method of selection enables the training of a potential which demonstrates good transferability \cite{Bartok2018}.
However, due to the nature of the sampling, it lacks the dense population of configurations around particular local minima which we find are important for achieving very high accuracy on particular crystalline properties.
We therefore choose to augment the training dataset selected through FPS with a number of mandatory configurations chosen using chemical intuition, focused on the bulk crystalline phases and certain defect and surface structures. 
Specifically, we note that optimised geometries for structures used in the validation sections of this paper are included in the training. 
The final database is comprised of the union of the 4000 FPS-selected points and the existing GAP-17 dataset, while a further ca. 1000 configurations are manually added to target specific properties.
The selected configurations, as well as a representation of their position in phase space, can be seen illustrated in figure (Fig. 1). 
This sketch-map \cite{Ceriotti2011,De2016} representation of the total training dataset uses the same measure of kernel similarity as discussed above to position points in a reduced dimensionality such that points which are similar in the full high-dimensional descriptor space are closer together, and those which are dissimilar are further apart. 


This sketch-map representation also serves as a qualitative overview of the type of structures to which we fit our model.
Structures with carbon atoms of highly varied coordination environments, from $\mathrm{sp}^{1}$ to $\mathrm{sp}^{2}$ and $\mathrm{sp}^{3}$ can be seen. 
Those allotropes which are $\mathrm{sp}^{2}$ hybridised, such as graphene, graphite and carbon nanotubes are clustered together towards the right of the map.
Amorphous structures can be seen as a large region in the centre, with low density ($\mathrm{sp}^{1}$ and $\mathrm{sp}^{2}$ rich) amorphous carbon at the far right, high density $\mathrm{sp}^{3}$ rich amorphous carbon towards the left, eventually approaching crystalline diamond at the very far left of the map. 
The more exotic, sometimes hypothetical structures collected from the SACADA database are often found separated from bulk crystalline or amorphous configurations.
In the far top right of the map, isolated gas phase dimer configurations are found.
\section{Training of the Potential}
%
%
%
%

\begin{figure}[H]
    \centering
    \includegraphics[width=0.5\textwidth]{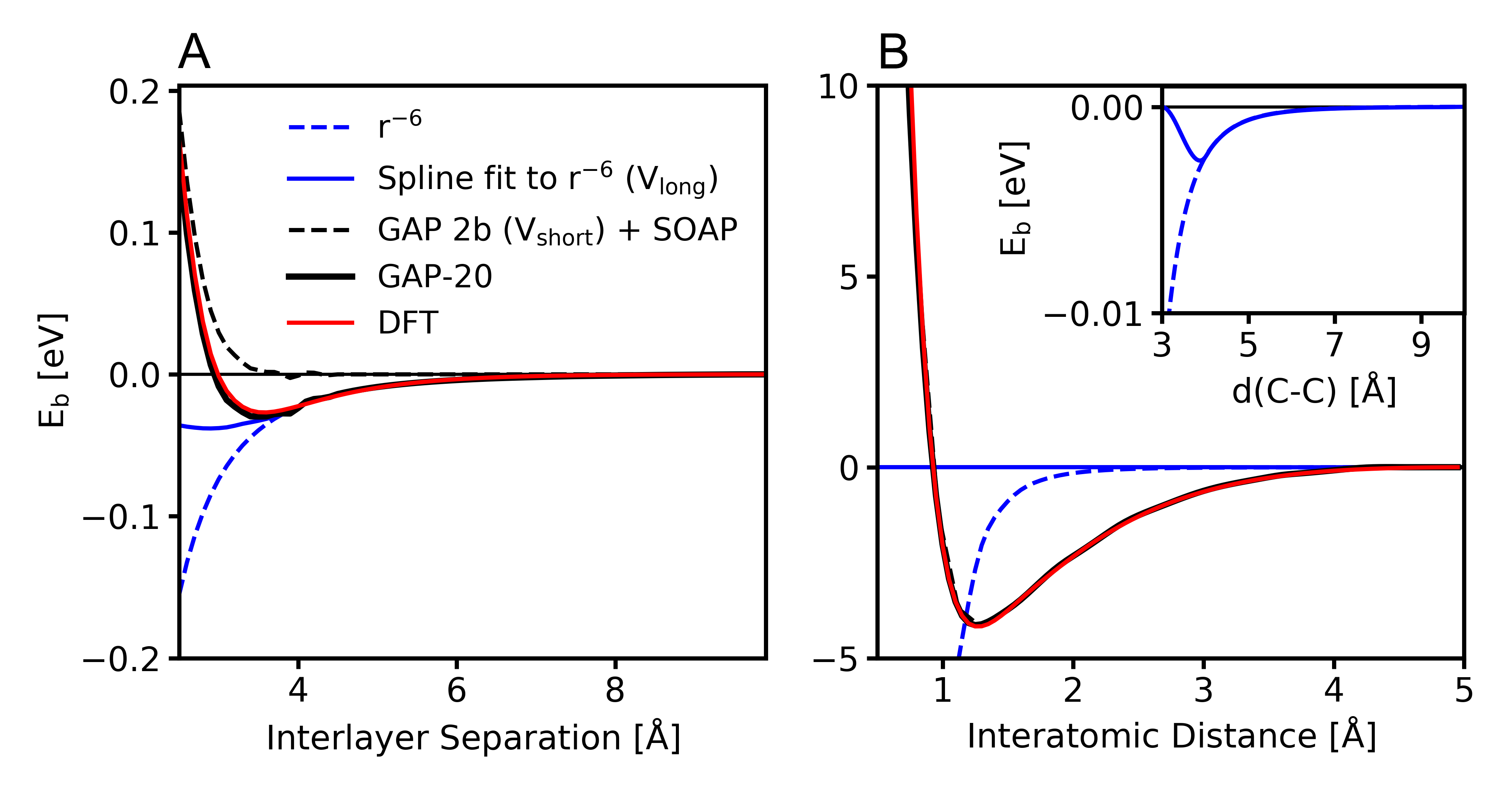}
    \caption{Construction of the long-range 2b component of the GAP model. An analytical spline is fitted to a function which decays as $\mathrm{r}^{-6}$, designed to recover the long-range attraction between graphene layers. (a) Shows the predicted energies for each component for the interaction of two graphene layers at different distances. The long range attraction is well characterised by the $\mathrm{r}^{-6}$ component of the $\mathrm{r}^{-6}$ potential, which is in turn well recovered by the analytical spline. The GAP fit using a 2b, ($\mathrm{V}_{\mathrm{short}}$), 3b and SOAP descriptor provides the appropriate repulsive potential at short distances but is too short ranged to describe the attractive tail. GAP-20 reproduces the whole curve with good accuracy across a range of distances. (b) Shows the same decomposition for the gas phase dimer. In this case, the strong bonding interactions are dominated by the GAP 2b ($\mathrm{V}_{\mathrm{short}}$), 3b and SOAP descriptor components. The energy of the $\mathrm{r}^{-6}$ component becomes large and negative for short distances. (b, inset) Provides a closer view of panel B, and shows how the 2b spline fit to the $\mathrm{r}^{-6}$ component is brought smoothly to 0 at a distance of 3 \AA.}
    \label{fig:new_two_body}
\end{figure}

%
We choose to construct GAP-20 to represent the PES as the combination of contributions from a two body (2b), a three body (3b) and a high dimensional many-body (MB) component.
It is an empirical observation that a large proportion of the interaction in an atomistic system may be satisfactorily captured by considering 2b interactions.
In particular this is the case for the exchange repulsion experienced as interatomic distances become very small. 
Representing this exchange repulsion in its full high-dimensional form would be costly from the perspective of both training data generation, potential generation and the ultimate evaluation of the potential. 
The nature of bonding interactions for carbon may also be captured in an approximate way, being generally attractive between 1.2 - 1.6 \AA, with an attractive tail at long distances. 
We design the 2b part of our model as a GAP fitted 2b component ($\mathrm{V}_{\mathrm{short}}$) r $<$ 4.0 \AA.
For larger separations (10 $\geq$ r $>$ 4.0 \AA), this smoothly transitions to an analytical spline potential ($\mathrm{V}_{\mathrm{long}}$) which decays as $\mathrm{r}^{-6}$.
This long range component is fitted to correctly reproduce (albeit without many-body contributions) the long-range attraction due to van der Waals interactions of graphitic layers. 
A smooth transition is achieved by first fitting the analytical form of $\mathrm{V}_{\mathrm{long}}$ to the graphene bilayer interaction curve from 3.0 to 10.0 \AA.
$\mathrm{V}_{\mathrm{short}}$ is then trained by first subtracting $\mathrm{V}_{\mathrm{long}}$ from the total energy and fitting to the difference.
The resultant 2b potential (Fig. \ref{fig:new_two_body}) simply has the final form $\mathrm{V}_{\mathrm{short}} + \mathrm{V}_{\mathrm{long}}$
The true subtleties of interatomic bonding are inherently many-body in character, however.
%
%
We represent these higher-order contributions to the potential energy using a combination of a 3b descriptor and the aforementioned SOAP descriptor. 
The full details of the construction of the 3b and SOAP descriptors is given in detail elsewhere \cite{Bartok2013a,Bartok2013,Dragoni2017,Deringer2017,Rowe2018,Fujikake2018,Szlachta2014}. 

In short, the 3b term is a symmetrized transformation of the Cartesian coordinates of triplets of atoms, which is designed to be permutationally invariant to the swapping the atomic indices \cite{Deringer2017,Rowe2018}.
In the construction of the SOAP descriptor, the local environment around a target atom is represented by a `local neighbour density', constructed by placing a Gaussian basis function on each neighbouring atom within a certain cutoff, which we choose to be 4.5 \AA.
The Gaussian basis functions are scaled by a factor of $1/\mathrm{r}^{0.5}$ to reflect the greater contribution to material properties of atoms which are closer together \cite{Willatt2018,Faber2018,Christensen2020}.
Other functional forms for the radial scaling exist and the introduction of this scaling was performed independently of the optimisation of the SOAP descriptor cutoff, the choice of which is motivated below.
As a result, there may still be scope for further optimisation of these parameter sets beyond what is performed here. 
The local neighbour density is expanded in a basis set of spherical harmonics, the coefficients of which form a `SOAP vector'.
In our case we use a basis set up to order l = 4 and n = 12, our motivation for which is discussed in the following paragraph. 
This SOAP vector constitutes a unique representation of the local environment, which satisfies the requirements of being translationally, rotationally and permutationally invariant.
The SOAP kernel, used for regression, is constructed as the scalar product of individual SOAP vectors.
Such a kernel is physically interpretable, as it corresponds directly to the integral of two neighbour densities for all possible 3D rotations.
Details for the specific choices for a number of associated hyper-parameters are given in the supplementary material, while further details on their significance is given elsewhere \cite{Bartok2010,Bartok2013a,Bartok2013,Dragoni2017,Deringer2017,Rowe2018,Fujikake2018,Szlachta2014}. 

\begin{figure}[H]
    \centering
    \includegraphics[width=0.45\textwidth]{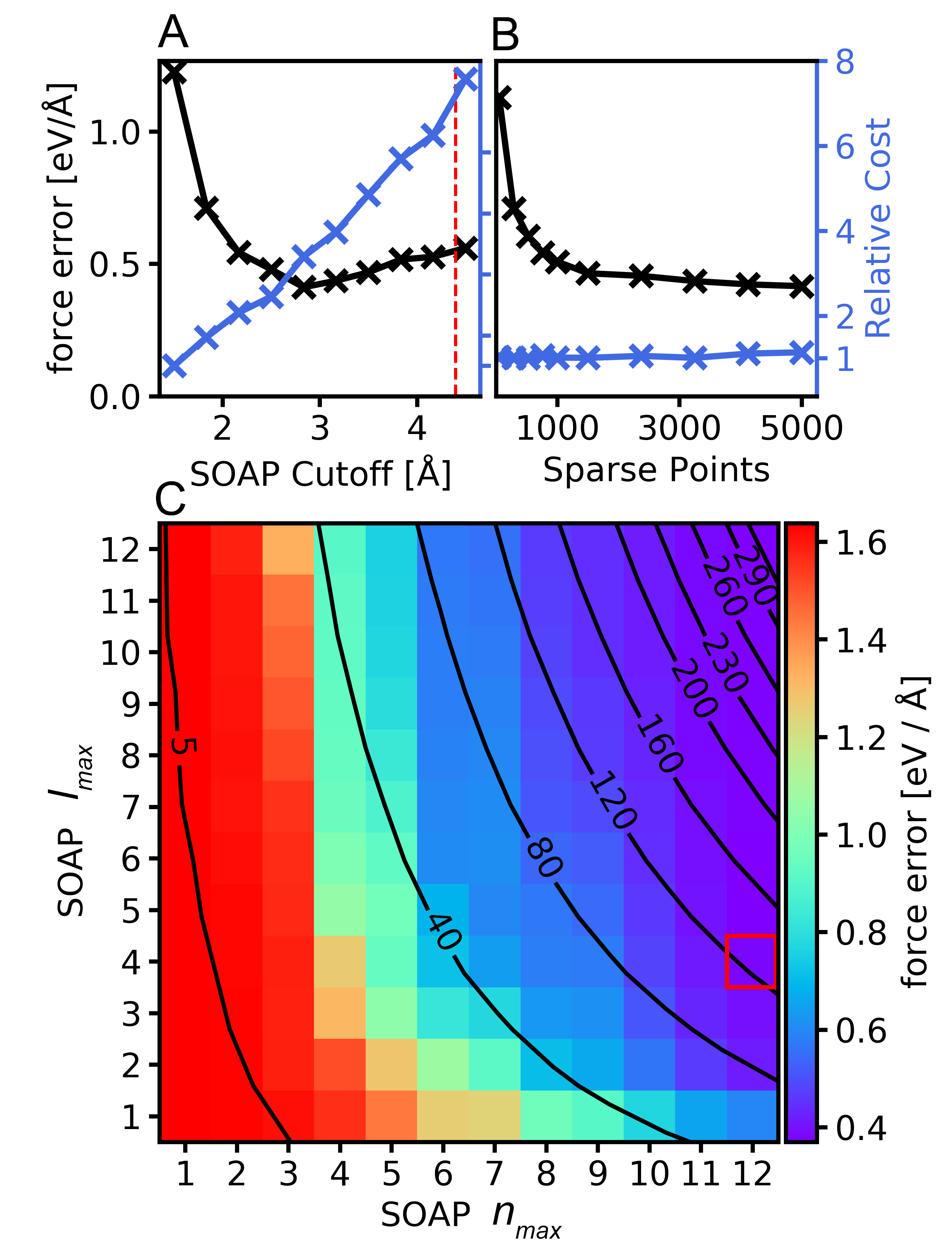}
    \caption{Mean absolute force errors calculated for an independent test set of configurations for different SOAP descriptors. (a) Force error behaviour and cost of evaluation (relative to the fastest GAP) of the resultant model as a function of the SOAP descriptor cutoff, the selected value of 4.5 \AA{} is indicated by the dashed red line. (b) Force error convergence and relative cost as a function of the number of sparse points included in the training. (c) Dependence of the force error and model evaluation cost on the order of the SOAP neighbour density basis set expansion. Force errors are indicated by the colour bar, while relative costs are shown by contour lines, our choice $l_{\rm max}$ = 4, $n_{\rm max}$ = 12 is highlighted by the red square.}
    \label{fig:optimisation}
\end{figure}

We now provide the details of select convergence tests for the optimisation of our GAP model.
These tests consider the independent optimisation of the SOAP descriptor cutoff, number of sparse points and the order of the radial basis set expansion.
In general we begin with a SOAP descriptor with an expansion of the neighbour density up to $l_{\rm max}$ = 8, $n_{\rm max}$ = 8, a cutoff of 4.2 \AA, $\sigma_{\mathrm{force}} = 0.01$ eV \AA$^{-1}$ and $\sigma_{\mathrm{energy}} = 0.001$ eV and $\zeta = 4$. 
We modify one parameter at a time in isolation, while keeping the remainder fixed.
We calculate the force error for the resulting models on a randomly selected independent set of test configurations which is not included in the training.
Figure \ref{fig:optimisation}(a)  shows the behaviour of the force errors as a function of the SOAP descriptor cutoff.
The force error has a minimum for a cutoff of 2.9 \AA, after which it begins to rise again as the increased size of the descriptor space expands beyond what can be populated with the number of available training configurations.
A na\"ive optimisation of these parameters based purely on the force errors would therefore select a cutoff radius of 2.9 \AA.
However, selection of these parameters cannot be performed in isolation from the intended application of the potential, but must also be motivated by knowledge of the behaviour of the material of interest.
In this regard, the force (or energy) error alone may be regarded as an imperfect or incomplete target property for optimisation.
Specifically, we find that although the minimum in the force error is found at much shorter distances, a longer cutoff of 4.5 \AA{} must be used in order to correctly describe graphitic structures, a feature which we consider to be mandatory for this potential.
The inter-layer spacing of graphitic structures is typically large (approx. 3.3 \AA) and a potential must incorporate enough of the structure of both layers to correctly model properties such as the binding curve of graphene layers or the energy difference between AA and AB stacked graphite.
The effect of these short cutoffs can be seen in the unsatisfactory behaviour of the Tersoff and REBO-II models when modelling the interlayer spacing of graphite (Table \ref{tab:crystalline_properties}), or graphene bilayers (Supplementary fig. 6). 
However, the problem of producing a single analytical metric for optimisation, which satisfactorily includes properties such as the lattice parameters, defect formation energies or phonon errors  as well as the force errors themselves is a challenging one.
In this instance, the design choice of selecting an appropriate descriptor cutoff remains partly qualitative in nature. 
Figure \ref{fig:optimisation}(b) shows the convergence of the mean absolute force errors as a function of the number of sparse points used in the training, this may be considered as a measure of the data saturation of GAP-20.
The force errors decrease rapidly up to approx. 1500 sparse points, at which point they begin to level off, although we note that a further increase in the number of sparse points has a negligible impact on the cost of evaluating the model.
Our choice of 9000 sparse points is therefore very tightly converged.
%

%
In figure \ref{fig:optimisation}(c) we show how the force error for our model converges as a function of the order of the basis set of radial functions used to expand the SOAP neighbour density.
The relative computational cost of each basis set is indicated on the same plot by labelled contour lines.
We find that the radial ($n$) component of the expansion, typically has a greater impact on the rate of convergence than the angular ($l$) component.
While previously in the construction of GAP models, band limits $n_{\rm max} = l_{\rm max}$, were used for the SOAP descriptor, we find that surprisingly, an improvement in accuracy can be achieved with essentially no additional cost by making a selection for the basis set expansion which is strongly biased to the radial ($n_{\rm max}$) component.
Of course the cost must also be taken into account, the use of a larger radial component is more expensive than an identical increase in the angular component, due to the greater number of basis functions introduced. 
Our selection of $l_{\rm max}$ = 4, $n_{\rm max}$ = 12 is chosen as a compromise between accuracy and efficiency.
Although a small improvement in the force errors can be achieved by selecting $n_{\rm max}= 12, \, l_{\rm max} > 4$, the resultant increase in the cost of training restricts both the size of the training data set which can be used and the size and length scales to which the resultant potential can be applied. 
\section{Crystalline Carbon}

\begin{figure}[H]
    \centering
    \includegraphics[width=0.4\textwidth]{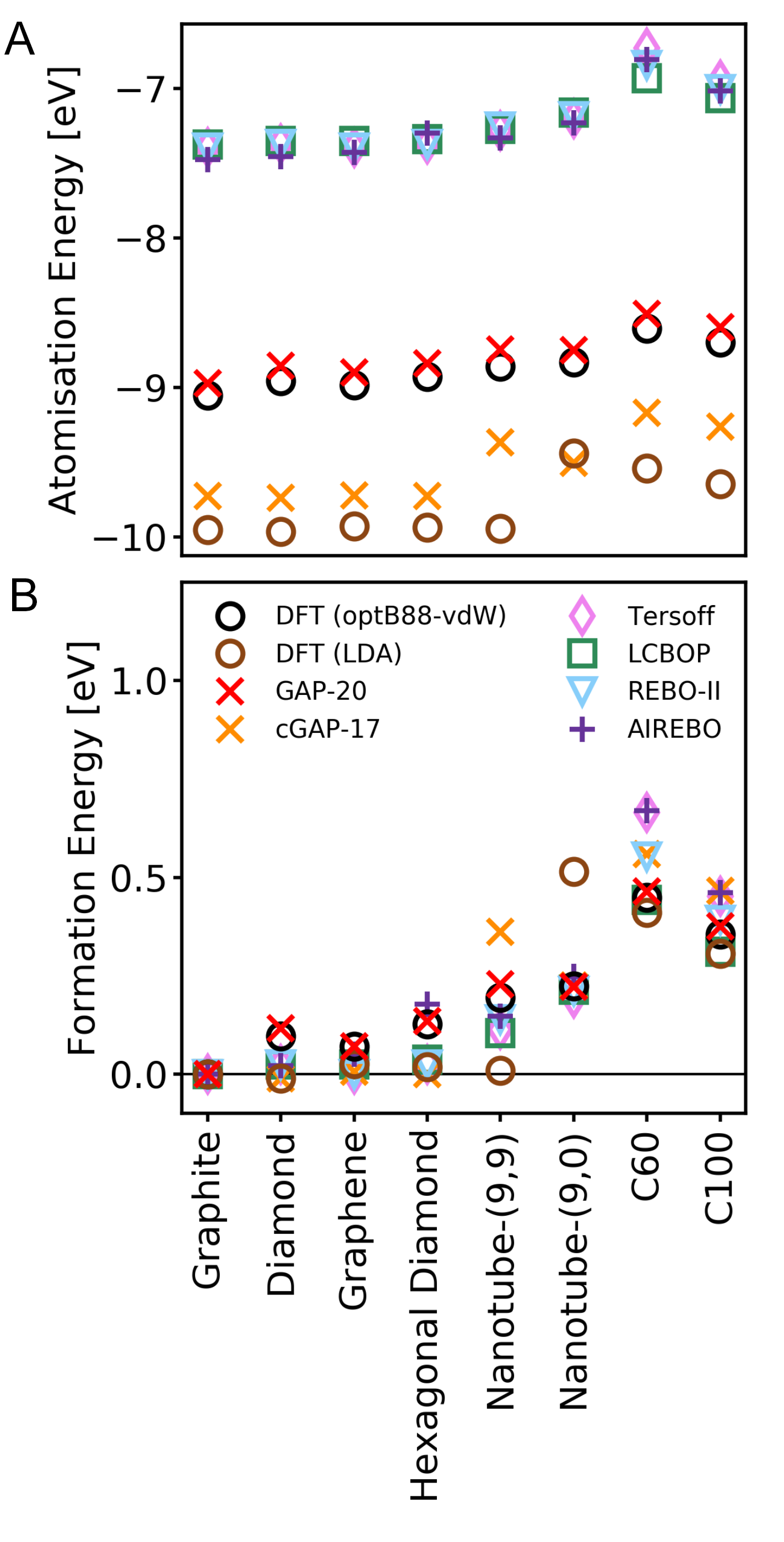}
    \caption{Formation energies of the crystalline phases of carbon, comparing results from DFT (optB88-vdW and LDA) to those from GAP-20 and the other models tested. (a) Atomisation energies using the isolated gas phase carbon atom as a reference, differences are dominated by overstabilisation of the gas phase atom by empirical models. (b) Formation energies given relative to the graphite formation energy of each particular model.}
    \label{fig:formation_energies}
\end{figure}

%
Among the most important material properties for any potential to predict accurately are those of the bulk crystalline phases.
Table 1 compares the lattice parameters and bond lengths as predicted by GAP-20 to those from the reference DFT method, in addition to a number of empirical models.
In figure \ref{fig:formation_energies}, we also provide both the atomisation energies, and the formation energies of the crystalline phases relative to the thermodynamically stable state of carbon (at standard temperature and pressure), i.e. graphite.
We define the atomisation energy of a phase relative to the isolated gas phase carbon atom $\mathrm{E}_{\mathrm{at}}$ as, 

\begin{equation}
    \mathrm{E}_{\mathrm{f}} = \mathrm{E}_{\mathrm{bulk}} - n \mathrm{E}_{\mathrm{at}},
\end{equation}

\noindent
where $E_{\mathrm{bulk}}$ is the energy of the bulk phase and n is the number of atoms in the bulk.
Lattice parameters are calculated by independently optimising the cell vectors for each allotrope, until the total energy is converged to less than $10^{-4}$ eV.
GAP-20 accurately predicts the lattice parameters and bond lengths of all of the tested crystalline allotropes with an average error of 0.2\%, and their formation energy to within 0.5\%. 
Accurately modelling the graphite c lattice parameter, corresponding to the spacing between graphitic layers proved particularly challenging for candidate GAP models, as did the formation energy.
This is in large part due to the shallow nature of the energetic minimum characterising the graphite inter-layer interactions and the long range of the atomic descriptor required to capture it.
As discussed above, the choice of SOAP cut-off was specifically informed by a desire to capture this property correctly.
We consider this in particular to be a mandatory characteristic of a general carbon potential which would be absent for any model with a shorter cut-off.
It is useful here to make a brief comparison to selected empirical potentials. 
While we do include DFT reference data for all properties, in subsequent sections these reference values are only computed using the same level of DFT used to train GAP-20.
For benchmarking GAP-20, which is our primary purpose, this is not problematic, however we do not fully account for the potential impact of functional dependence, or the errors of DFT with respect to experiment, when making comparisons to empirical models.
Many of the empirical models considered are fitted to experimental data, or contain values from other DFT functionals, typically LDA, which should be taken into account when comparing different model predictions to our optB88-vdW reference values.
To give some indication of the functional dependence, reference values for the formation energies in figure \ref{fig:formation_energies} are given using both the optB88-vdW and LDA functionals.
We also re-iterate that the GAP-17 model was fitted to LDA data, so it would be expected to accurately reproduce DFT values at this level only.
On average, GAP-20 predicts the lattice parameters of the tested crystalline phases with an error of 0.2\%, while the Tersoff, LCBOP, REBO-II, AIREBO potentials have errors averaging 5\%, 0.3\%, 4\% and 1\% respectively (Table \ref{tab:crystalline_properties}. 
What the Tersoff and REBO-II potentials gain in efficiency by using short cutoffs, they lose in accuracy, notably by predicting dramatically incorrect inter-layer spacings (c lattice parameters) for graphite. 
This error is fixed by the inclusion of medium and long-range terms to account for van der Waals interactions in the LCBOP and AIREBO models, however.
Despite its inaccuracy for graphene, the REBO-II potential does achieve good accuracy on the remaining lattice parameters; the additional terms included in the bond-order potential constitute a dramatic improvement over the Tersoff potential. 
Due in part to its complete reparameterisation to account for the effects of long-range interactions in the bond-order part of the potential, LCBOP does outperform the other empirical potentials tested here in most cases.

\begin{table*}[t]
    \caption{Lattice parameters and bond lengths of the crystalline carbon phases. In the case of fullerenes, bond lengths are given in lieu of lattice parameters. Absolute values for the lattice parameters are given, with percentage errors relative to DFT in brackets. The Tersoff and REBO-II potentials have no interaction between graphite layers for any physically reasonable lattice parameters and as such these values are omitted.}
    \begin{adjustbox}{center} 
        \begin{tabular}{lcccccc}
             \toprule
             & \multicolumn{6}{c}{Lattice Parameter(s) [\AA] (\% Error)} \\ 
             \cmidrule(r){2-7}
             & {DFT} & {GAP-20} & {Tersoff} & {LCBOP} & {REBO-II} & {AIREBO} \\
             \midrule
             Graphite (\textit{a})       & 2.46  & 2.47 (0.4)  & 2.53 (2.4)  & 2.46 (0.4)  & 2.46 (0.4) & 2.42 (2.0) \\   
             
             Graphite (\textit{c})       & 6.65  & 6.71 (0.9)  & -           & 6.36 (4.4)  & -          & 6.72 (1.1) \\
             
             Graphene           & 2.46  & 2.46 (0.0)  & 2.53 (2.8)  & 2.46 (0.0)  & 2.46 (0.0) & 2.42 (1.6) \\                  
             
             Diamond            & 3.58  & 3.59 (0.3)  & 3.57 (0.3)  & 3.57 (0.3)  & 3.57 (0.3) & 3.56 (0.6) \\            
             
             Hexagonal Diamond        & 2.52  & 2.53 (0.4)  & 2.52 (0.0)  & 2.52 (0.0)  & 2.52 (0.0) & 2.52 (0.0) \\                    
             
             Nanotube-(9, 9)    & 4.26  & 4.25 (0.2)  & 4.35 (2.1)  & 4.24 (0.5)  & 4.26 (0.0) & 4.18 (1.9) \\                    
             
             Nanotube-(9, 0)    & 2.41  & 2.39 (0.8)  & 2.53 (5.0)  & 2.47 (2.5)  & 2.47 (2.5) & 2.43 (0.8) \\                    
             
             $\mathrm{C}_{60}$ Fullerene      & 1.40  & 1.40 (0.0)  & 1.46 (4.5)  & 1.41 (0.7)  & 1.42 (1.4) & 1.40 (0.0) \\               
             
             $\mathrm{C}_{100}$ Fullerene     & 1.39  & 1.39 (0.0)  & 1.39 (0.0)  & 1.39 (0.0)  & 1.39 (0.0) & 1.39 (0.0) \\         
             \bottomrule
        \end{tabular}
    \end{adjustbox}
    \label{tab:crystalline_properties}
\end{table*}

In absolute terms, the atomisation energies (fig. \ref{fig:formation_energies}(a)) from the tested empirical potentials differ significantly from those predicted by both reference DFT methods, due to the very different energies of the isolated gas phase atom.
In the case of GAP-17, the small offset between the LDA reference and the model prediction is the result of the isolated atom not being included in the training dataset. 
When using the formation energy of graphite as a reference state however, (fig. \ref{fig:formation_energies}(b)) this offset is removed and the agreement between the empirical models and DFT improves considerably. 
When using both the gas phase atom and graphite as a reference, there is an excellent agreement between GAP-20 and the optB88-vdW DFT reference for all of the phases considered here.
GAP-20 uniformly predicts the atomisation energies of the tested allotropes to within an error of 1\%, including the relatively subtle difference in energetics between normal cubic and hexagonal diamond and the energetics of nanotubes and fullerenes.
The inclusion of the gas-phase atom in the training is vital to accurately predict these atomisation energies. 
There is surprisingly little difference between the formation energies predicted by the different many-body potentials tested here, though there are a few points of note. 
Firstly, due to their short cutoffs, the Tersoff and REBO-II potentials do not distinguish between graphite and graphene as the thermodynamically stable phase and as such their formation energies are predicted to be equal.
Similarly, only the GAP-20, LCBOP and AIREBO models correctly favour cubic over hexagonal diamond, although the AIREBO model overestimates the difference in energy by a factor of 5, while the other models considered do not distinguish between the two diamond phases. 
A more complete evaluation of the formation energies for different chiralities of nanotubes is given in the supplemental material, for GAP-20, the energy errors for a significantly wider range of structure types are also given. 

In addition to the static properties of the crystalline allotropes, it is an important characteristic of any potential that it accurately model the lattice dynamics of bulk crystals, i.e. their behaviour at finite temperature. 
The phonon spectrum of a material is a direct probe of this which is experimentally measurable.
In addition, a number of thermodynamically relevant properties, including the thermal expansion coefficient and the constant volume heat capacity of a material may be calculated directly from the phonon dispersion relation by calculation of the free energy.
It is clear therefore, why a correct prediction of the phonon dispersion relation is a highly desirable feature of an interatomic potential.

\begin{figure}[H]
    \centering
    \includegraphics[width=0.4\textwidth]{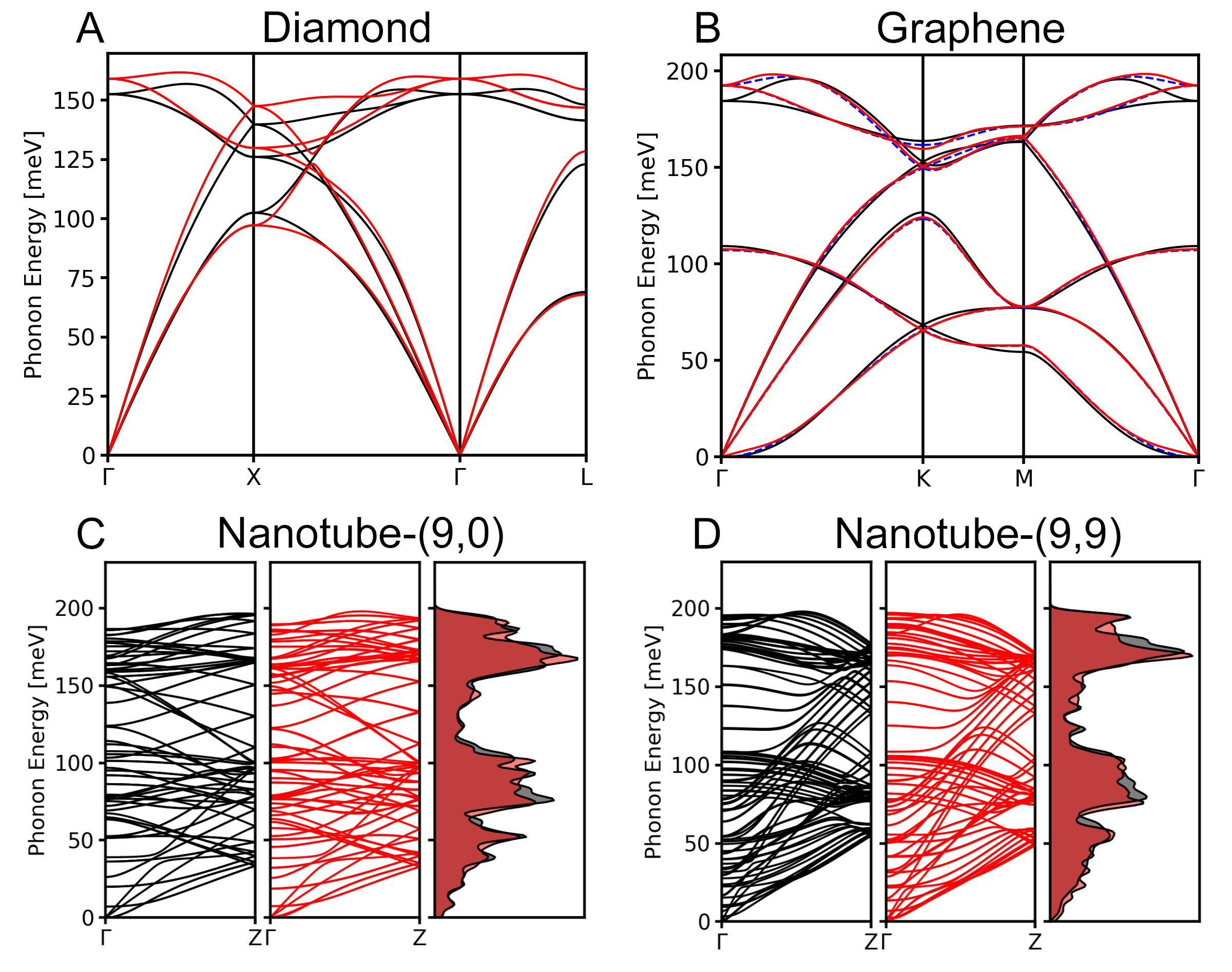}
    \caption{A. Phonon dispersion relation for diamond as predicted by GAP-20 (black) with comparison to DFT (optB88-vdW) reference data (red). B. Graphene phonon dispersion relation comparing GAP-20 and DFT (optB88-vdW) reference data. The dashed blue line shows the predicted phonon dispersion curve for the graphene-only model previously published \cite{Rowe2018} C. (9,0)-Nanotube phonon dispersion and vibrational density of states. D. (9, 9)-Nanotube phonon dispersion relation and vibrational density of states. Equivalent comparisons for the other models tested are given in the supplementary material.}
    \label{fig:gr_dia_phonons}
\end{figure}

Figure \ref{fig:gr_dia_phonons} shows the comparison between the phonon dispersion curves calculated using the reference DFT method and those calculated using the carbon GAP model for graphene, diamond, a zig-zag (9, 9) and an armchair (9, 0) carbon nanotube. 
Phonon dispersion curves were computed using the finite displacement method as implemented in the Phonopy Python package \cite{Tanaka2015}.
Equivalent curves for the other models tested are provided in the supplementary material. 
We have previously reported results comparing the phonon dispersion relation for a purely graphene GAP model, to those from experimental x-ray scattering data and a number of reference DFT methods \cite{Rowe2018}.
It is useful to make a comparison between the highly targeted model previously published and the much more general GAP-20 introduced here. 
A particular concern might be that significantly expanding the configurational space on which we wish to train, as we have done here, would necessarily damage the quality of the predictions for graphene compared to the previous model – particularly for a property as sensitive as the phonon dispersion curves.
It is demonstrated in figure \ref{fig:gr_dia_phonons}(b) that this is not the case; the dispersion relation of the phonon curves for graphene from GAP-20 are comparable to those of the previously published graphene GAP model \cite{Rowe2018}.
The energies of the phonon bands are correctly predicted across all of the high symmetry directions plotted, while the frequencies (in particular at the high symmetry points) are found to be correct to within 4 meV, which may be compared to a value of 1 meV for the pristine graphene model \cite{Rowe2018}. 
The quality of the GAP-20 model prediction is comparable for diamond (which cannot be modelled at all with the pristine graphene model), though with marginally larger errors for the prediction of the energies of certain bands, up to 7 meV for the higher frequency modes. 
GAP-20 also captures the difference in vibrational behaviour between armchair and zig-zag nanotubes remarkably well. 
There are some differences in the energy of certain splittings for some bands, but the magnitude of these errors is small, typically on the order of a 2-3 meV. 
In particular, it can be seen from fig. \ref{fig:gr_dia_phonons} that the vibrational density of states for the two nanotube systems agrees well with the DFT reference. 
\section{Surfaces of Carbon}
From the point of view of atomistic simulation, surfaces present a major challenge, as their correct description requires a treatment of a number of competing physical interactions \cite{Ristein2006,Kern1997,Kern1998a,Kern1998,Ooi2006}. 
We compute the surface energy for each model by first optimising the bulk structure for the parent crystal until the total energy is converged to $10^{-3}$ eV. 
We then cut the surface along the desired direction and compute the specific surface energy $\gamma_{s}$ at T = 0 K as,

\begin{equation}
  \gamma_{\mathrm{s}} = (\mathrm{E}_{n} - \mathrm{nE}_{\mathrm{bulk}}) / \mathrm{2A},
\end{equation}

\noindent
where $\mathrm{E}_{\mathrm{n}}$ is the energy of the n slab layer containing two surfaces, which may be as-cut (unrelaxed) or allowed to relax and $\mathrm{E}_{\mathrm{bulk}}$ is the energy of a single atom in the bulk structure and A is the area of the surface structure. 
In the case of the amorphous surfaces, due to the extent of the surface relaxation observed, we report only the as-cut surface energies.
Graphite may be readily cleaved to expose its (0001) surface, which is remarkably stable and is by far the predominant face of graphite, while in diamond, the (100), (111) and (110) surfaces are of particular interest \cite{Thinius2016}. 
We also compute the as-cut surface energies for an ensemble of amorphous structures, by cutting bulk amorphous systems along different directions.

\begin{table*}[t]
    \caption{Surface energies of low Miller index surfaces for common carbon allotropes. Reference energies are calculated using DFT, absolute values from each model are given, with their percentage error in brackets. Note that for the amorphous surfaces, the surface energy is averaged over a large number of different surfaces. In the amorphous case, the error provided is the average of the individual point-wise errors, rather than the error between the average surface energies.}
    \begin{adjustbox}{center} 
        \begin{tabular}{lcccccc}
             \toprule
             & \multicolumn{6}{c}{Surface Energy [eV  \AA$^ {-2}$] (\% Error)} \\ 
             \cmidrule(r){2-7}
             & {DFT} & {GAP-20} & {Tersoff} & {LCBOP} & {REBO-II} & {AIREBO}\\
             
             \midrule
             Diamond (100) (As cut)                             & 0.56  & 0.60 (7) & 0.47 (16) & 0.61 (9) & 0.69 (23) & 0.73 (30) \\
             Diamond (100) (Relaxed)                            & 0.54  & 0.56 (4) & 0.42 (22) & 0.59 (9) & 0.69 (28) & 0.72 (33) \\
             \cmidrule(r){1-1}
             Diamond (111) (As cut)                             & 0.64  & 0.73 (14) & 0.88 (38) & 1.07 (67) & 1.00 (56) & 1.03 (61) \\
             Diamond (111) (Relaxed)                            & 0.62  & 0.66 (6) & 0.88 (42) & 1.07 (73) & 1.00 (61) & 1.03 (66) \\
             \cmidrule(r){1-1}
             Diamond (110) (As cut)                             & 0.68  & 0.70 (3) & 0.70 (3) & 0.89 (31) & 0.74 (9) & 0.74 (9) \\
             Diamond (110) (Relaxed)                            & 0.68  & 0.67 (1) & 0.63 (7) & 0.83 (22) & 0.69 (1) & 0.69 (1) \\
             \midrule
             Graphite ($0001$)  (As Cut)                        & 0.015  & 0.013 (13)  & 0 (100) & 0.005 (67) & 0 (100) & 0.011 (27) \\
             Graphite ($0001$)  (Relaxed)                       & 0.015  & 0.012 (20)  & 0 (100) & 0.005 (67) & 0 (100) & 0.011 (27) \\
             \midrule
             Amorphous Surfaces (As Cut)                      & 0.26  & 0.27 (4) & 0.25 (4)  & 0.25 (4) & 0.25 (4)  & 0.25 (4) \\
             \bottomrule
        \end{tabular}
    \end{adjustbox}
    \label{tab:surfaces}
\end{table*}

The energies of several important surface cuts and their reconstructions are given in Table \ref{tab:surfaces}.
GAP-20 typically predicts the diamond surface energies correctly to within 7 \%, the exception being the case of the relaxed diamond (111) surface, where the error is slightly larger at 15 \%. 
The structures of the relaxed surfaces were also found to be in excellent agreement, with the average error in the positions of individual surface atoms being $~10^{-3}$ \AA.
The graphite (0001) surface energy is extremely small and it thus proved challenging to produce a model which could correctly characterise this, however, GAP-20 predicts the unrelaxed and relaxed surface energies correctly to within an error of 3 meV \AA$^{-2}$ (20 \%). 
With the inclusion of vdW interactions considered in their construction, the LCBOP and AIREBO potentials both predict the graphite (0001) surface energy rather well, with errors of 67 and 27 $\%$ respectively.

While GAP-20 achieves low errors for the surface energies of all the diamond surfaces considered, the other models generally perform well for at least one diamond surface, though none exhibit uniformly low errors.
The Tersoff, REBO-II and AIREBO models predict the energies of the diamond (110) surfaces to within 10 \% of the reference value.
Conversely, of the empirical models only the LCBOP potential correctly predicts the energy of the diamond (100) surface; errors for the Tersoff, REBO-II and AIREBO potentials were 22, 28 and 33\% respectively. 
None of the empirical potentials performed well for the (111) surface of diamond.
The Tersoff and REBO-II models do not show any binding between graphitic layers for any reasonable initial geometry. 
This would lead to the spontaneous exfoliation of graphite layers and the eventual disintegration of graphite crystals in simulations employing these models.

\section{Defective Carbon}
A certain concentration of defects is a guarantee in any experimental material sample. 
Such imperfections may have a strong impact on the structural, optical and thermal properties of a material and may be introduced into a crystal structure to induce or modify properties.
The engineering of defects is of great technological importance and consequently their accurate modelling by an interatomic potential is highly desirable.
The possibility of rehybridisation, which allows carbon atoms to reconstruct with differing numbers of bonds to stabilise particular structures allows carbon to have a wider variety of defects than most other elements.
To the best of our knowledge, there is not a set of defect formation energies for a wide range of carbon defects computed at precisely the same level of theory. 
Therefore, we here assemble such a reference set, for which we compute defect formation energies in large supercells to avoid defect self-interaction in the computation of energies. 
For graphite, a $(6 \times 6 \times 2)$ supercell with 288 atoms and four graphite layers was used \cite{Li2005,Probert2003}.
In the case of graphene, a $(10 \times 10)$ supercell with 200 atoms was employed and for diamond a $(3 \times 3 \times 3)$ supercell with 216 atoms was used \cite{Xu2018,Rowe2018}.
Defect formation energies are calculated for the representative (9, 9) and (9, 0) index carbon nanotubes, which had 174 and 180 atoms in the supercells used respectively \cite{Charlier2002}. 
For each structure, the lattice parameters and ionic positions of the pristine structures were optimised as discussed previously.
The ionic positions of the defective structures were then optimised until the energy was converged to within $10^{-5}$ eV, while keeping the lattice parameters fixed. 
We compute the formation energy $\mathrm{E}_{\mathrm{f}}$ of a vacancy defect relative to the energy of an atom in an ideal parent structure:

\begin{equation}
    E_{\mathrm{f}} =  E_{\mathrm{d}} - (\mathrm{n}E_{\mathrm{at}} + E_{\mathrm{bulk}})
\end{equation}

\noindent
Where $\mathrm{E}_{\mathrm{d}}$ is the energy of the defective supercell structure, $\mathrm{E}_{\mathrm{bulk}}$ is the energy of the undefective bulk structure and $E_{\mathrm{at}}$ is the energy of a single atom in the bulk structure, while n is the number of carbon atoms added (positive n) or removed (negative n) to form the defect. 
The simplest of defects involves the absence of one or two atoms from their regular position in the lattice, forming monovacancy and divacancy defects.
Monovacancy defects often result in unsaturated bonds at the defect site, while divacancy structures, particularly in $\mathrm{sp}^{2}$ hybridised systems, can reconstruct to produce saturated configurations.
In graphene, graphite and carbon nanotubes, the 14-membered ring formed by the removal of two adjacent atoms from the structure reconstructs to form a saturated $\mathrm{sp}^{2}$ structure with two 5-membered and one 8-membered ring - a more stable structure known as a 5-8-5 divacancy.
In graphene, this defect may further reconstruct to remove the unfavourable 8-membered ring to form a 555-777 or 5555-6-7777 divacancy reconstruction. 
Monovacancy coalescence is also observed in diamond, whereupon annealing at high temperature, monovacancies migrate to form divacancy defects, with fewer unsaturated bonds per absent carbon atom. 

\begin{figure}[H]
    \centering
    \includegraphics[width=0.45\textwidth]{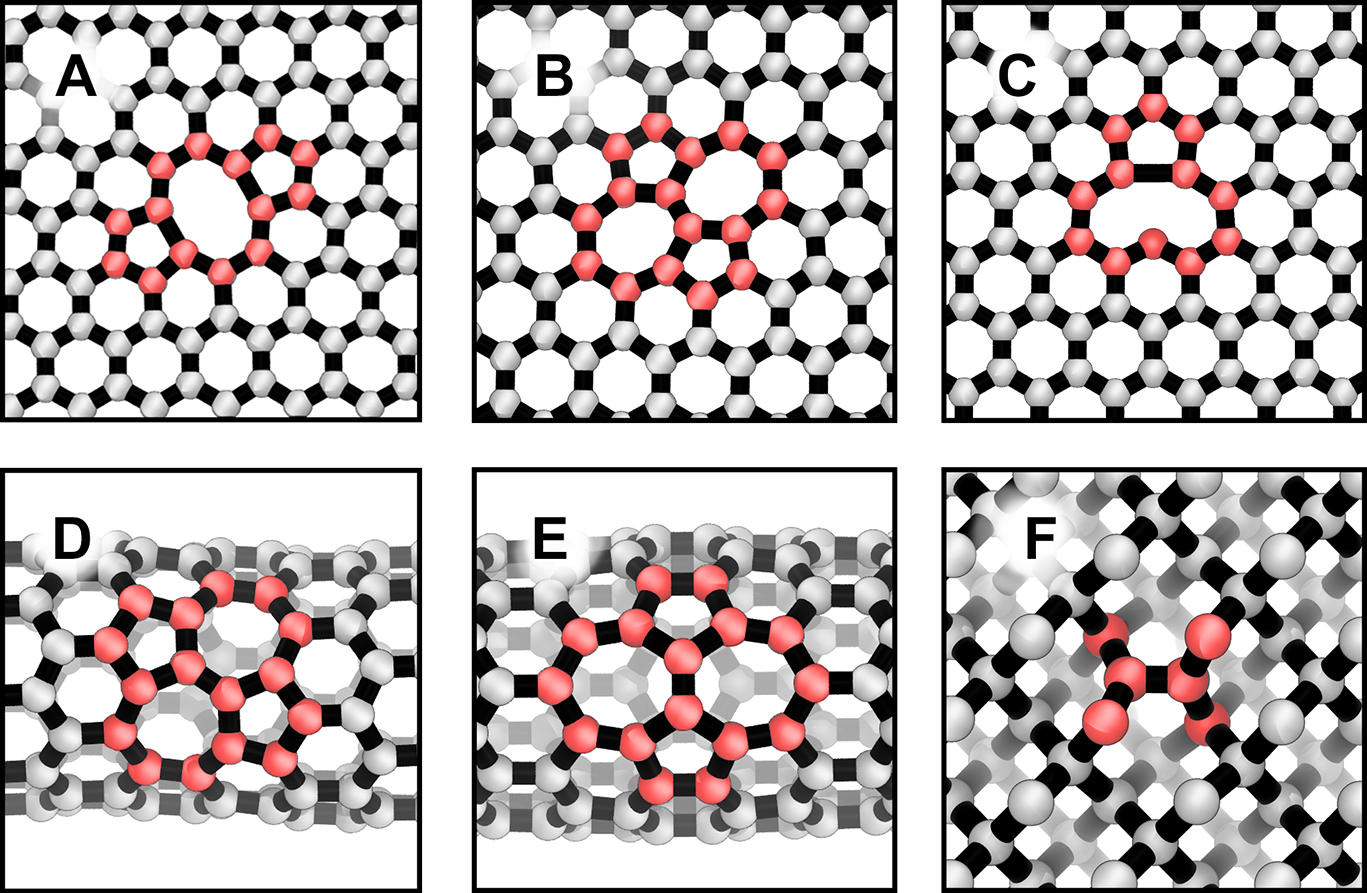}
    \caption{Images of selected carbon defect structures, with atoms in the immediate vicinity of the defect highlighted in red. (a) graphene divacancy defect (b) graphene Stone-Wales defect (c) graphene monovacancy (d) (9, 9)-nanotube Stone-Wales defect (transverse orientation) (e) (9, 9)-nanotube Stone-Wales defect (parallel orientation) (f) diamond split interstitial defect}
    \label{fig:defect_images}
\end{figure}

Graphite is the only allotrope of carbon in which true interstitial defects are known, wherein interstitial atoms may be found between graphite layers \cite{Li2005}. 
The most stable arrangement of this is in a ‘dumbbell’ configuration, where the adatom displaces an atom in the graphite structure to form a symmetric arrangement of trigonally bonded carbon atoms above and below the sheet.
Isolated interstitial atoms are not known either experimentally or from theory to be stable in diamond, rather a split interstitial is found, where a lattice site is shared by two carbon atoms which are displaced along the [100] and $[\bar{1}00]$ directions \cite{Hunt2000}. 
In $\mathrm{sp}^{2}$ bonded allotropes of carbon, the rotation of a single C-C bond transforms four 6-membered rings into a cluster of two 7-membered and two 5-membered rings, forming a Stone-Wales type defect \cite{Stone1986,Kotakoski2011,Ma2009,Kotakoski2011}.

\begin{table*}[t]
    \caption{Formation energies of common defects in carbon structures for GAP-20 and the other models considered, with DFT (optB88-vdW) values given as reference. Data are given in eV, with percentage errors relative to DFT given in brackets. In each case, the value given is for the optimal geometry of the defect found with that particular model.}
    \begin{adjustbox}{center} 
        \begin{tabular}{lccccccc}
             \toprule
             & \multicolumn{6}{c}{Formation Energy [eV] (\% Error)} \\ 
             \cmidrule(r){2-7} & \\
             & {DFT} & {GAP-20} & {Tersoff} & {LCBOP} & {REBO-II} & {AIREBO} \\
             
             \midrule
             Graphene Stone-Wales                     & 4.9 & 4.8 (2) & 1.9 (61) & 4.5 (8)  & 5.3 (8) & 5.4 (10) \\
             Graphene Monovacancy                     & 7.7 & 7.0 (8)  & 2.5 (68) & 6.9 (10) & 7.5 (3)  & 7.2 (6) \\
             Graphene Divacancy (5-8-5)               & 7.4 & 7.9 (7)  & 5.1 (31) & 7.5 (1) & 7.5 (1) & 9.2 (24) \\
             Graphene Divacancy (555-777)             & 6.6 & 6.9 (5) & 5.2 (21) & 6.6 (0)  & 6.8 (3)  & 8.7 (32)\\
             Graphene Divacancy (5555-6-7777)         & 6.9 & 7.4 (7) & 7.9 (14) & 7.2 (4)  & 7.6 (10) & 9.5 (38) \\
             Graphene Adatom                          & 6.4 & 5.9 (8) & 6.7 (5)  & 6.8 (6) & 7.4 (16)  &  7.8 (22) \\
             
             \midrule
             Graphite Monovacancy                     & 7.8 & 7.3 (6) & 7.1 (9)  & 7.8 (0) & 7.9 (1) & 7.6 (3)\\
             Graphite Divacancy (5-8-5)               & 9.6 & 9.2 (4) & 12.6 (31)  & 8.2 (15) & 8.0 (17) & 9.7 (1)\\
             Graphite Stone-Wales                     & 5.4 & 5.6 (4) & 12.8 (137) & 5.7 (6)  & 6.0 (11) & 6.0 (11)\\
             Graphite Interstitial                    & 7.4 & 7.9 (7)  & 9.7 (31)  & 7.2 (3) & 7.1 (4)  &  6.8 (8) \\

             \midrule
             Diamond Monovacancy                      & 6.6 & 4.3 (35)  & 5.2 (36) & 7.2 (11)  & 7.1 (4)  & 6.8 (8)\\
             Diamond Divacancy                        & 9.1  & 6.6 (27)  & 5.1 (44) & 10.6 (16) & 10.7 (18) & 10.1 (16)\\
             Diamond Split Interstitial               & 11.4  & 8.3 (27)  & 12.4 (9)  & 9.8 (14) & 11.0 (4)  & 11.4 (0)\\ 
             
             \midrule
             Nanotube-(9, 9) Monovacancy              & 6.4  & 5.8 (9) & -5.1 (180) & 3.8 (41) & -1.6 (125) & -2.5 (139)\\
             Nanotube-(9, 9) Divacancy                & 4.7  & 4.8 (2)  & -5.5 (217) & 2.9 (38) & -0.9 (119) & -2.3 (149)\\
             Nanotube-(9, 9) Stone-Wales (Parallel)   & 4.4  & 4.5 (2)  & -4.5 (202) & 2.1 (52) & -2.1 (148) & -3.9 (189)\\
             Nanotube-(9, 9) Stone-Wales (Transverse) & 3.5  & 3.5 (0) & -5.8 (261) & 2.0 (44) & -3.3 (192) & -2.00 (156)\\
             
             \midrule
             Nanotube-(9, 0) Monovacancy              & 5.3  & 4.9 (8) & -0.9 (117) & 4.4 (17) & 4.7 (11) & 3.4 (36)\\
             Nanotube-(9, 0) Divacancy                & 3.6  & 3.5 (3)  & -1.0 (128) & 3.0 (17) & 4.1 (14) & 2.8 (22)\\
             Nanotube-(9, 0) Stone-Wales (Parallel)   & 2.7  & 3.1 (15)  & -1.3 (148) & 3.2 (19) & 3.4 (26) & 3.6 (33)\\
             Nanotube-(9, 0) Stone-Wales (Transverse) & 3.5  & 3.2 (9)  & -1.1 (131) & 3.1 (11) & 4.2 (20) & 2.6 (26)\\
             
             \bottomrule
        \end{tabular}
    \end{adjustbox}
    \label{tab:defect_properties}
\end{table*}

Table \ref{tab:defect_properties} compares the energies of a number of defects as computed with DFT, GAP-20 and the other models considered.
In most cases, GAP-20 correctly predicts the defect formation energy to within an error of 10\%. 
Typically, the prediction of the formation energies of Stone-Wales type defects was found to be extremely accurate, with no error (to within the precision of the values given) in either the graphite or graphene cases and only small errors for nanotubes.
The errors for the formation energies of diamond defects tend to be larger, ranging from 25-35\%, while those for defective nanotubes range from 0-11\%.
Anecdotally, we note that although relevant training data for the defects considered are represented in the training data, it proved challenging to achieve defect formation energies which were universally accurate.
In particular this is due to the sensitivity of the formation energies to aspects such as the SOAP descriptor cutoff, specific training data included and the number of sparse points used in the training.
Considering the empirical potentials, we find that the modifications to the Tersoff potential included in the REBO-II model dramatically improve the quality of the predicted defect formation energies; percentage errors are often decreased by an order of magnitude or more when comparing these two potentials.
Surprisingly, these results show that the inclusion of the long-range Lennard-Jones term in the AIREBO model often has a negative impact on the accuracy of its predicted defect formation energies, indicating that the addition of a long-range term without reparameterisation of the short-range components has adversely impacted the energetics of the model.
Indeed, in the case of LCBOP, where this reparameterisation of the short range bond-order potential has been performed, we find that the errors are significantly reduced, and are in many cases comparable with the performance of GAP-20.
The exception to this being the case of defective nanotubes, where LCBOP exhibits errors ranging from 11-52\%. 
In fact, the prediction of nanotube defect formation energies proved challenging for all of the empirical models considered.
In a number of cases, defect formation was found to be an energetically favourable process and was associated with a strong relaxation of the nanotube structure after defects were induced. 

As well as accurately predicting the energetic cost of inducing defects in carbon structures, GAP-20 was also found to accurately predict the structures of these defects. 
We quantify this accuracy by calculating the structural similarity between the defect structures optimised with our GAP model and those from DFT, in the form of the root mean squared error (RMSE) between the two optimally overlapped structures. 
In all but a handful of cases, the RMSE for these defects is vanishingly small, with atoms having an error in their position of less than $10^{-2}$ \AA, when comparing identical atoms from GAP-20 and DFT structures.
In particular, the presence and height of the characteristic buckling of the Stone-Wales defect in graphene was well described, as was the structural distortion resulting from the presence of defects in both (9,9) and (9,0) index carbon nanotubes.
Similarly, the rehybridisation and reconstruction of (5-8-5), (555-777) and (5555-6-7777) graphene divacancy defects was accurately reproduced, as were the geometries of all of the diamond defects considered.
Situations in which GAP-20 showed structural inaccuracies were the nanotube monovacancy structures and the parallel Stone-Wales defect in the (9,9) index nanotube, for which the GAP model predicted a larger distortion of the bulk nanotube structure due to the presence of the defects. 
We also find, that as with all the models considered here, GAP-20 does not correctly describe the asymmetry introduced through a Jahn-Teller distortion of the graphene monovacancy defect – instead predicting the monovacancy to have a symmetric geometry. 
This is perhaps unsurprising as the energy difference between the symmetric and asymmetric geometries is typically small (ca 350 meV). 
However, even in the cases illustrated here the typical error in the position of any atom was found to be only 0.1 \AA.
That GAP-20 is capable of accurately modelling both the energetics and structural characteristics of a wide range of carbon defects indicates its potential usefulness in a wide range of simulations in which defective structures may be relevant, including fracture, atom bombardment and simulations of membrane characteristics. 
\section{Liquid Carbon}
\begin{figure}[H]
    \centering
    \includegraphics[width=0.3\textwidth]{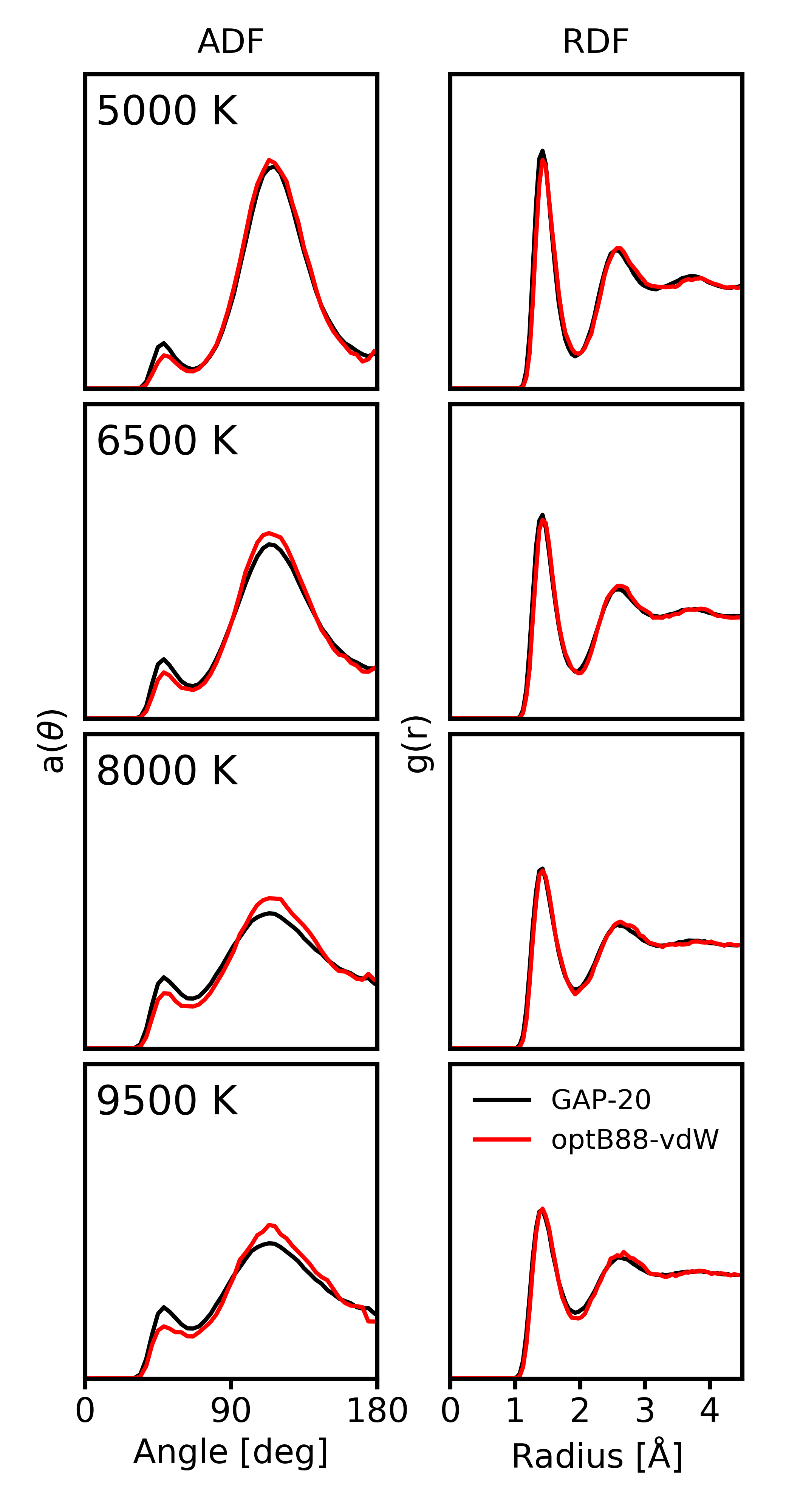}
    \caption{Angular and radial distribution functions for liquid carbon at a fixed density of 2.5 $\mathrm{g \, cm}^{-3}$ for temperatures between 5000 - 9500 K. GAP-20 results are shown in black, while reference DFT (optB88-vdW) data are given in red.}
    \label{fig:liquid_temperature}
\end{figure}

\begin{figure}[H]
    \centering
    \includegraphics[width=0.3\textwidth]{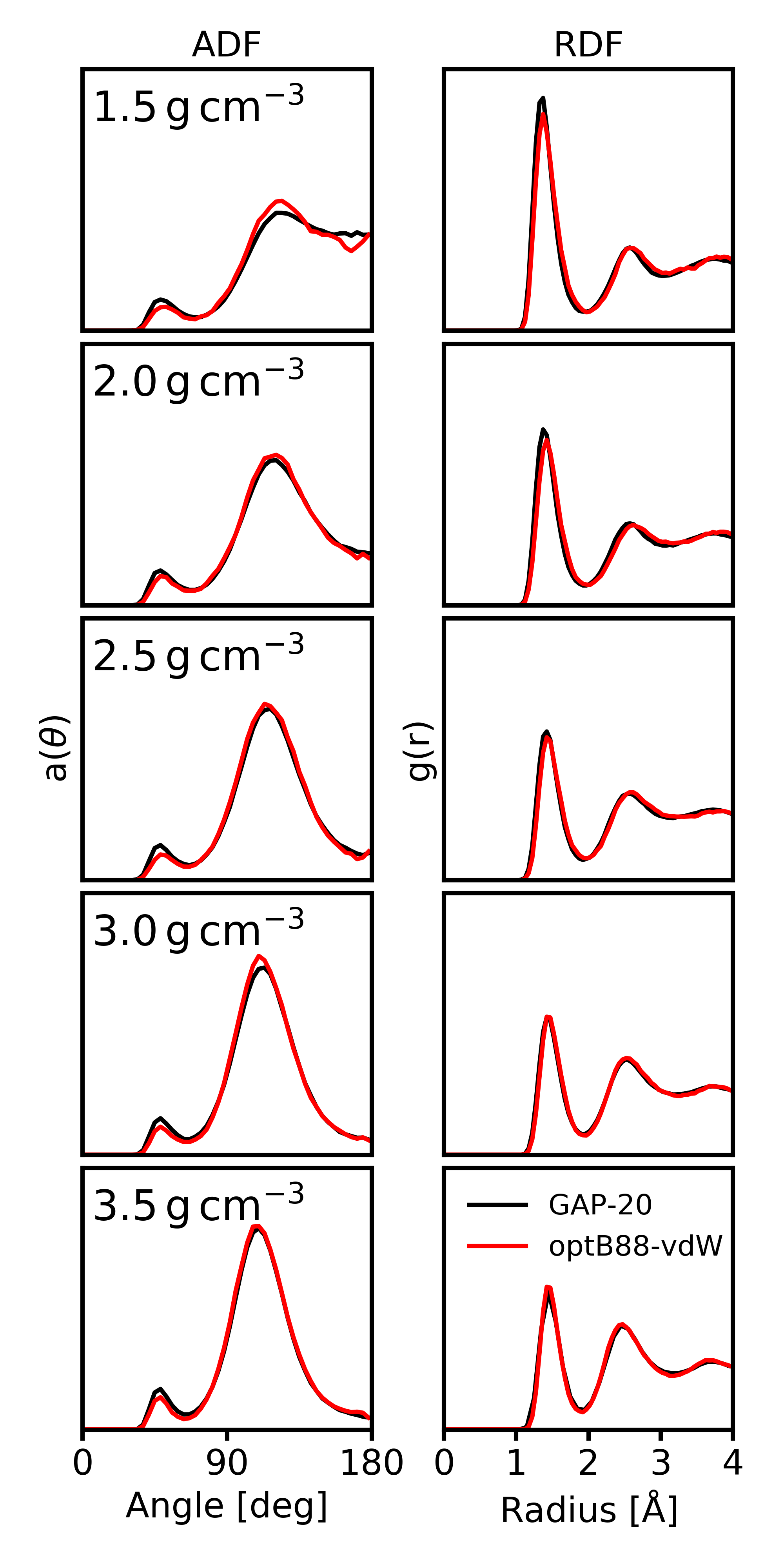}
    \caption{Angular and radial distribution functions for liquid carbon at 5000 K for a range of densities from 1.5 to 3.5 $\textrm{g \, cm}^{-3}$. GAP-20 results are shown in black, while reference DFT (optB88-vdW) data are given in red.}
    \label{fig:liquid_density}
\end{figure}

As discussed previously, the requirements of a potential for the satisfactory modelling of crystalline and liquid or amorphous phases are significantly different.
In the case of crystalline materials, a highly accurate description close to a local minimum for a system is required \cite{Rowe2018}. 
Conversely, in a liquid simulation, a vastly greater number of local configurations are explored, requiring a high degree of flexibility and transferability \cite{Deringer2017}.
As in the case of GAP-17, we therefore use the liquid as a benchmark for the flexibility of our potential \cite{Deringer2017}, scanning over a wide range of densities and (here) temperatures. 
The aim is to diagnose any possible issues which might be exposed by visiting a very diverse set of configurations during the simulations.
There is a strong precedent for the study of high temperature liquids, including carbon, using DFT \cite{Correa2006,Pozzo2013}.
A good agreement with DFT-MD data is therefore strong evidence for the usefulness of the potential for further studies of liquid carbon, which is present only under extreme conditions, but is nonetheless vitally important, e.g. for understanding the nucleation and formation of diamond and graphite under a wide range of circumstances \cite{Strong1970,Sorkin2006,Gust1980,Mundy2008}. 
The radial distribution function (RDF) of a liquid represents a convenient measure of its local structuring, as does its angular distribution function (ADF).
Here, we compare the results of constant volume \textit{ab initio} molecular dynamics simulations to those of GAP-20.
We perform two sets of simulations, one for a range of densities between $1.5 - 3.5 \, \mathrm{g \, cm}^{-3}$ at 5000 K and the other for a range of temperatures between 5000 - 9500 K at a fixed density of 2.5 $\mathrm{g \, cm}^{-3}$.
These simulations were performed for 216 atom systems using a chain of 5 Nos\'e-Hoover thermostats.
\textit{Ab initio} trajectories were generated using VASP, simulations were performed at the gamma point and data were collected for 5 ps at each temperature and pressure \cite{Kresse1996,Kresse1996a,Kresse1999}.
We find that there is a very good agreement between the \textit{ab initio} data and the GAP-20 predictions for both the RDF and ADF across the wide range of temperatures studied (see figure \ref{fig:liquid_temperature}).
GAP-20 correctly models the increased structuring of the liquid carbon as the temperature is reduced from 9500-4500 K.
At temperatures below approximately 3500 K, the GAP model predicts the liquid to form an amorphous glass which slowly graphitises (which is entirely expected because the temperature is now below the melting line).
While a full discussion on the mechanism of formation and resulting morphology of graphitised amorphous carbons generated using GAP-20 is beyond the scope of the current work, this process has previously been shown to be an excellent method of differentiation between the numerous available carbon potentials \cite{Tomas2016,DeTomas2019}.
Figure \ref{fig:liquid_density} shows the RDF and ADF computed with both GAP-20 and optB88-vdW across a wide range of densities, from $1.5$ to $3.5 \, \mathrm{g \, cm}^{-3}$ at 5000 K.
This test is particularly important as it represents dynamical simulations of structures from highly $\mathrm{sp}^{1}$ and $\mathrm{sp}^{2}$ hybridised (low density) through to a predominantly $\mathrm{sp}^{3}$ hybridised liquid at higher density.
GAP-20 captures this change in the bonding characteristics of liquid carbon, in particular the increase in the $\mathrm{sp}^{1}$ hybridised fraction of the liquid at very low densities (reflected in bond angles close to 180 degrees), qualitatively similar to GAP-17 \cite{Deringer2017}.

That GAP-20 can model the atomistic structure of liquid carbon at a wide variety of temperatures and densities, while maintaining the ability to accurately predict properties such as the phonon relation and defect formation energies is a reflection of the flexibility of the GAP methodology. 
Such wildly different systems explore a range of characteristic energies, where important fluctuations cover many orders of magnitude; in carbon, this can be anywhere from the meV range in the case of differences between graphite defect energies to fluctuations on the order of tens of electron volts as encountered in the liquid.
Despite these very different energy ranges, it is not unlikely that a potential may encounter all of them over the course of a single simulation (for example during the crystallisation of a solid phase directly from the liquid) and it is therefore important that they be handled correctly.

\section{Transferability of the Potential}
Ultimately, the purpose of any interatomic potential is that it may be used for the discovery of new and interesting phenomena. 
Consequently,in its application it may encounter structures which were not explicitly considered in its construction, in this case meaning that it must model structures which were not included in the training data base.
It has therefore been a criticism of ML potentials that their poor performance in extrapolation might inhibit their use for scientific discovery.
As discussed earlier, the problem of extrapolation is circumvented by the fact that we consider only the local environment around a particular atom to be important for predicting its atomic energy and the forces acting upon it.
While the problem of exploring the entirety of the 3N dimensional chemical space is indeed intractable, sufficiently sampling all of the physically relevant local environments is not \cite{Bartok2018}. 

We demonstrate this here by performing a diagnostic GAP driven random structure search (GAP-RSS), similar in spirit to Refs. \cite{Deringer2018c} and \cite{Deringer2017a}, and demonstrate that the predicted energies of these structures agree well with those from DFT \cite{Pickard2011a,Deringer2018,Deringer2017a}. 
We then calculate a number of high energy pathways for specific transformations not included in the training and compare these to DFT.
Both of these tests serve the purpose of exploring the high energy regions of the potential energy surface which may be explored during molecular dynamics simulations or geometry optimisation and which must be well described for an ML model to be transferable.
Importantly, they are both explicitly designed to include configurations which are not present in the training data set of GAP-20. 
To perform the first test, we generate a cubic unit cell with lattice parameter a = 3 \AA. 
In this cell, we randomly place 8 carbon atoms, avoiding any overlaps such that the distance between any two carbon atoms is not less than 2 \AA. 
This process is performed to generate 1000 initial randomised geometries. 
The LAMMPS package is then used to optimise lattice vectors of the cell independently using conjugate gradient descent, while maintaining their orthogonality, until the total energy is converged to within $10^{-8}$ eV \cite{Plimpton1995}.
Following this, the positions of the atoms in the unit cell are optimised using a FIRE algorithm \cite{Bitzek2006}, until the total energy is again converged to within $10^{-8}$ eV.
This cycle is repeated twice more before performing a final conjugate gradient optimisation of the atomic positions and cell vectors until the total energy is converged to $10^{-10}$ eV. 
\begin{figure}[H]
    \centering
    \includegraphics[width=0.5\textwidth]{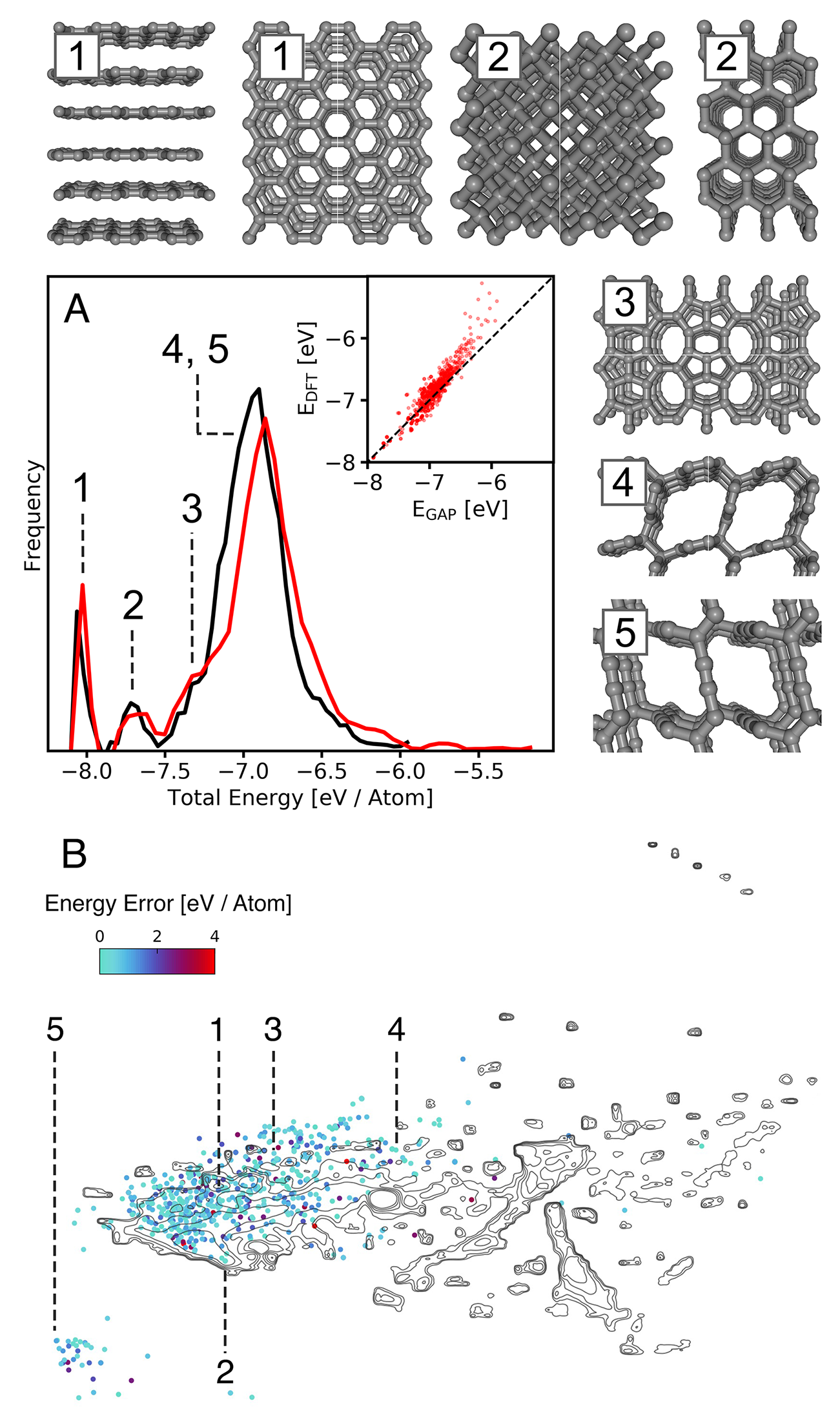}
    \caption{(a) A comparison of the histograms of energies of structures identified by GAP-RSS, given in eV / atom, showing good agreement for the prediction of the energy of structures between GAP-20 (shown in black) and DFT (optB88-vdW) (shown in red). A number of examples of structures identified in GAP-20 driven random structure search are shown. The position of each of the example structures on the histogram is indicated by their numbering, 1) AB stacked graphite, AA stacked graphite. 2) cubic and hexagonal diamond. 3) Haekelite 4) crosslinked graphitic structure. 5) Novel carbon structure with high proportion of $\mathrm{sp}^1$ hybridised carbon atoms (b) The structures resulting from the GAP-RSS projected into the sketch-map representation from Fig \ref{fig:sketchmap}. The density of the structures present in the training data are indicated by the contour lines, while the structures identified from the GAP-RSS are shown as individual points.}
    \label{fig:RSS}
\end{figure}

To validate the results of our GAP-RSS, we recompute the energies of the structures found using the reference DFT method used to train the model.
We note that for across all 1000 structures, the predicted energy agrees well with the energy predicted from DFT. 
It has previously been shown that correctly identifying low energy structures from a RSS is an extremely challenging task for empirical models, which often predict qualitatively incorrect behaviour and fail to find physically relevant configurations due to their having many more local minima than the DFT PES \cite{Pickard2011,Needs2016}.
Our GAP-RSS correctly identifies a range of important low-energy carbon allotropes, as well as numerous more exotic species. 
In particular, AB-stacked graphite was found as the lowest energy allotrope of carbon. 
AA- and ABC- stacked graphite allotropes are also identified in the search, their energy is correctly predicted to be higher than that of the AB stacked graphite structure. 
Furthermore, both diamond and lonsdaleite are both correctly identified. 
We also identify a number of more exotic carbon allotropes, some of which are known either from experiment or theory but were not included in the training dataseset, including crosslinked graphite structures, porous carbon cages and a variety of haekelite structures.
For the vast majority of structures found during the GAP-20 driven random structure search, the predicted energies from both DFT and GAP-20 agree well (figure \ref{fig:RSS}).
We also return at this stage to the sketch-map representation of the training dataset given in figure \ref{fig:sketchmap}. 
In figure \ref{fig:RSS}(b) we provide a projection of the GAP-RSS structures onto this sketch-map representation.
GAP-RSS points are coloured according to the GAP-20 energy error. 
The density of structures present in the original training dataset is indicated by black contour lines.  
It is clear that most structures found are clustered in the region representing the bulk amorphous and crystalline polymorphs, with very few structures representative of fullerenes or nanotubes identified. 
This is a reflection of the fact that only 8 atoms are included in the unit cell used for the RSS.
Additionally, the RSS procedure employed begins with simulation cells which are fully periodic and with no symmetry constraints imposed on the initial atomic positions. 
In the lower left of the sketch-map is a cluster which is structurally distinct from those present in the training data, as indicated by its large separation from other points in the sketch-map. 
These structures are characterised by their highly $\mathrm{sp}^1$ rich character.
Although a significant number of amorphous structures which are rich in $\mathrm{sp}^1$ hybridised carbon atoms are included in the training data, there are indeed very few crystalline $\mathrm{sp}^1$ rich structures. 
Despite being structurally distinct from anything included in the training dataset, the error in the GAP-20 prediction for the energy of these structures remains low.
This indicates excellent performance for GAP-20 in applications where transferability to potentially novel structures is important. 

\begin{figure}[H]
    \centering
    \includegraphics[width=0.3\textwidth]{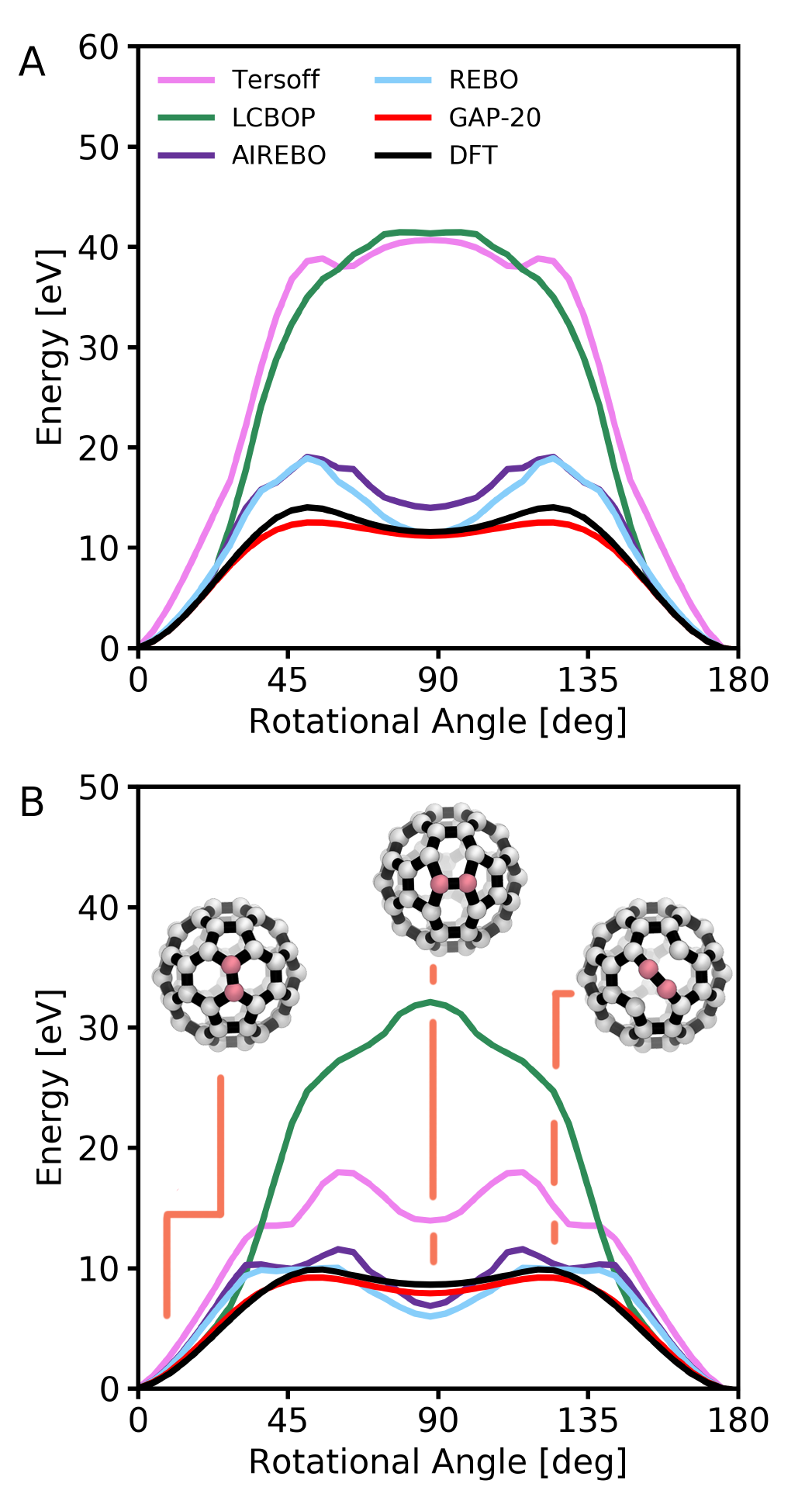}
    \caption{Energies for rigid transformations of a C-C bond in graphene (a) and in a C60 fullerene (b). Results from GAP-20, DFT (optB88-vdW) and a selection of empirical potentials are shown.}
    \label{fig:rigid_barriers}
\end{figure}

We also test GAP-20 on a number of specific structural transformations.
Although our GAP model is not trained explicitly on reaction barriers, it is useful to test how well the model performs for the prediction of the types of barriers which might be encountered in studies of the reactivity of carbon nanostructures. 
To this end, we compare the predictions of our GAP model to those of DFT for two approximate transformations; a rigid bond rotation in graphene and a C60 fullerene.
Since these calculations are performed on rigid structures, rather than for example using nudged elastic band calculations, the barriers calculated here will not be true defect formation barriers. 
They are, however, still representations of physically reasonable points on the potential energy surface which are not included in the training dataset and so form a useful test of the potential compared to other models \cite{Wales1987,Kumeda2003}.
Figure \ref{fig:rigid_barriers} (a) shows the barrier to rigid rotation of a C-C bond within a graphene sheet as predicted by GAP-20, and the tested empirical potentials. 
The performance of GAP-20 on this test is reassuring for its wider application, it achieves excellent accuracy with respect to both the height of the local minimum in the rotation and the height of the barrier.
The AIREBO and REBO potentials capture the general shape and height of the barrier, but predict jagged curves for the rotation, as compared to the smooth variation from DFT. 
The LCBOP and Tersoff potentials perform poorly, overestimating the energy of the rotation by more than 30 eV, and erroneously situating a maximum in the potential energy where a minimum is found from DFT. 
In the case of the Tersoff potential, two additional spurious minima are located close to where the DFT maxima are located. 
A similar situation is observed for the behaviour of the C60 rotational barrier in figure \ref{fig:rigid_barriers} (b). 
Here, it is again seen that GAP-20 performs well, providing a good estimation of the barrier height and shape with respect to DFT. 
The REBO, AIREBO and Tersoff potentials all situate spurious minima in the potential energy close to where the maxima in the reference DFT curve are located, although they do also predict a minimum at the correct rotation.
LCBOP again overestimates the energy of the barrier by 20 eV, and situates a maximum in the potential energy surface where a minimum ought to be located.
\section{Conclusion}
The advantages conferred by the flexibility of the Gaussian approximation potential framework are made clear by the wide variety of structures which are accurately treated by GAP-20. 
The variable hybridisation of carbon makes it an extremely challenging element to model using empirical potentials; its structurally diverse allotropes are energetically similar and the properties of these depend on an broad range of physical interactions, from the weak van der Waals forces binding graphite to the stiff covalent bonds of diamond.
We have demonstrated here a model which is equally suited to modelling not just these two bulk structures, but defects, surfaces and liquid carbon as well. 
Wherever possible, we have validated the performance of GAP-20 against the reference DFT method and shown it to perform well for a number of physical properties across the different phases.
Included in these are a number of processes involving bond breaking and formation, some of which have been challenging for cheaper empirical potentials by construction. 
Tests for transferability, specifically by diagnostic GAP-RSS runs and the study of transformations not included in the training, suggest that GAP-20 could readily be applied to more thorough explorations of the carbon potential energy landscape, for example, in the search for larger fullerenes \cite{Wales2004, Wales1999} or in crystal-structure prediction by expanding on Refs. \cite{Deringer2017a} and \cite{Deringer2018c}. 
Further applications may include the more detailed study of non-graphitising or ``hard" carbons \cite{Harris1997,Harris2005,Harris2014,Dou2019}, following on from earlier GAP-17 based studies in Ref. \cite{Deringer2018a} and \cite{Huang2019}.

Despite the many potential applications of GAP-20, the model is not without its shortcomings.
%
While it remains significantly more computationally affordable than direct \textit{ab initio} simulation (in particular for large systems) the cost of its evaluation is much greater than that of empirical potentials (Supplementary Fig. 12), and therefore the latter will still give access to even larger-scale systems \cite{DeTomas2019}. 
We also note that ‘real’ carbon is rarely found in isolation – hydrogenation and oxidation of carbon structures is not considered here. 
The expansion of the scope of the potential to treat hydrogenated or oxidised structures would complicate the process of training both by requiring a larger training dataset and by requiring the inclusion of a number of interactions not considered here.
In addition to long ranged van der Waals interactions (which are only considered approximately in the current work), the introduction of other elements introduces the associated complexities of substantial charge rearrangements: polar bonds and partial charges. 
Long ranged Coulomb interactions, dipole-dipole and higher order multipole interactions remain a challenge for ML potentials. 
We note that the combination of pure carbon simulations (using GAP-17) and subsequent density-functional analyses of hydrogenation and oxidation \cite{DeTomas2019} or metal intercalation \cite{Huang2019} has proven fruitful, and we expect that further studies of this type will be facilitated by GAP-20, particularly when low-density, dispersion-dominated nanostructures are concerned.

We believe that we have achieved an excellent compromise for our potential, in that it accurately models the wide range of structures required to make it broadly applicable.
We do not claim perfect accuracy for all properties, however; we accept that fitting to such a wide range of structures will necessarily impact the accuracy in some areas. 
Notably, a large number of structures which were generated as part of the total training dataset are excluded from the final training.
Conversely, many have been included which might be irrelevant for a researcher’s intended purpose.
With this in mind, we have made freely available the total training dataset (structures, energies, forces and virial coefficients) produced as part of this work.
While we do not believe that this will typically be necessary, it is a further virtue of the GAP framework that a potential may be readily retrained to suit a particular purpose simply by modifying the composition of the training configurations used, we believe it is beneficial to offer the opportunity for users to tune the model to target higher accuracy in a particular region of interest.
In addition to the training dataset, the potential introduced here is provided in the form of an XML file and has been made freely available, along with the GAP code at \url{http://www.libatoms.org}, it has the unique identifier GAP\_2020\_4\_27\_60\_2\_50\_5\_436 and may be used within the QUIP software package which can be found at \url{https://github.com/libAtoms/QUIP}.
GAP-20 may be used for simulations directly in LAMMPS, using the QUIP for LAMMPS plugin \cite{Plimpton1995}.
\section{Supplementary Material}
See [Supplementary Material] for full details of computed formation energies and phonon dispersion curves for all models, further information on GAP hyper-parameter selection and command line argument used, graphene bilayer separation curves and force errors for various configurations. More information on the optimisation and computational cost of GAP-20 compared to DFT is also given. 

\section{Acknowledgements}
A.M. was supported by the European Research Council under the European Union’s Seventh Framework Programme (FP/2007-2013) / ERC Grant Agreement No. 616121 (HeteroIce project). V.L.D. acknowledges a Leverhulme Early Career Fellowship and support from the Isaac Newton Trust. We are grateful to the UK Materials and Molecular Modelling Hub for computational resources, which is partially funded by the EPSRC (EP/P020194/1). We are also grateful for computational support from the UK national high-performance computing service, ARCHER, for which access was obtained via the UKCP consortium and funded by EPSRC grant ref EP/P022561/1. In addition, we are grateful for the use of the UCL Grace High Performance Computing Facility (Grace@UCL), and associated support services, in the completion of this work.

\section{Data Availability}
The data that support the findings of this study are openly available at \url{http://www.libatoms.org/Home/DataRepository} and on  request from the authors. 

\newpage

\bibliographystyle{ieeetr}


\end{document}


\maketitle

\newpage
 
In this supplemental material, we include data on further tests performed on GAP-20, in addition to further data for comparison with the other models considered in this work. In section 1, we give details of the GAP model parameters used in this text. In section 2, we provide tabulated data for the crystalline formation energies presented in the main text, this is followed in section 3 by further information on the predicted formation energies of a range of armchair and zig-zag nanotubes. Sections 4, 5, 6 and 7 give the phonon dispersion curves for graphene, diamond, a 9,9-index nanotube and a 9,0-index nanotube, for all of the models considered. In section 8, we plot the graphene bilayer separation curves for the models tested. In Section 9, the force and energy errors for a wide range of configurations considered in the training set are provided. In section 10, we give some further details on the optimisation of the GAP model. Finally, in section 11, we provide some information on the efficiency of the model compared to direct \textit{ab initio} simulation. 
 
\section{Training of the Potential}

\begin{table}[H]
    \caption{Hyperparameters of the GAP Model. Note that modified values for $\sigma_{\mathrm{energy}}$, $\sigma_{\mathrm{force}}$ and $\sigma_{\mathrm{virial}}$ are used for a number of sets of configurations.}
    \begin{adjustbox}{center} 
        \begin{tabular}{lc}
             \toprule
                GAP 2b Descriptor & \\
             \cmidrule{1-1}
                Cutoff \AA & 4.5 \\
                $\delta$ & 2.0 \\
                Sparse Method & uniform \\
                Covariance & Gaussian \\
                Sparse points & 15 \\
            \cmidrule{1-1}
                3b Descriptor & \\
            \cmidrule{1-1}
                Cutoff \AA & 2.5 \\
                $\delta$ & 0.05 \\
                Sparse Method & uniform \\
                Covariance & Gaussian \\
                Sparse points & 200 \\
            \cmidrule{1-1}
                SOAP Descriptor & \\
            \cmidrule{1-1}
                Cutoff \AA & 4.5 \\
                Cutoff width \AA & 1.0 \\
                $\delta$ & 0.2 \\
                Sparse method & CUR \\
                Sparse points & 9000 \\
                $l_{\mathrm{max}}$ & 4 \\
                $n_{\mathrm{max}}$ & 12 \\
                $\zeta$ & 4 \\
            \cmidrule{1-1}
                Global Parameters & \\
            \cmidrule{1-1}
                $\sigma_{\mathrm{energy}}$ [eV] & 0.001 \\
                $\sigma_{\mathrm{force}}$ [eV / \AA] & 0.1 \\
                $\sigma_{\mathrm{virial}}$ [eV] & 0.2 \\
             \bottomrule
        \end{tabular}
    \end{adjustbox}
    \label{tab:hyperparameters}
\end{table} 

\begin{table}[H]
    \centering
    \caption{Due to the differing energy scales involved in training, specific sets of configurations are fitted with customised values for $\sigma_{\mathrm{energy}}$, $\sigma_{\mathrm{force}}$ and $\sigma_{\mathrm{virial}}$. The values for these are tabulated here.}
    \begin{tabular}{c|c|c|c}
         \toprule
         Config Type &  $\sigma_{\mathrm{energy}}$ [eV] & $\sigma_{\mathrm{force}}$ [eV / \AA] & $\sigma_{\mathrm{virial}}$ [eV] \\
         \midrule
         Graphite & 0.001 & 0.01 & 0.05 \\
         Diamond & 0.001 & 0.01 & 0.05 \\
         Graphene & 0.001 & 0.01 & 0.05 \\
         Crystalline Bulk & 0.001 & 0.01 & 0.05 \\
         Nanotubes & 0.001 & 0.01 & 0.05 \\
         Fullerenes & 0.002 & 0.1 & 0.2 \\
         Amorphous Bulk & 0.005 & 0.2 & 0.2 \\
         Liquid & 0.050 & 0.5 & 0.5 \\
         Defects & 0.001 & 0.01 & 0.05 \\
         Surfaces & 0.002 & 0.1 & 0.2 \\
         Amorphous Surfaces & 0.005 & 0.2 & 0.2 \\
         Liquid Interface & 0.050 & 0.5 & 0.5 \\
         Graphite Layer Separation & 0.001 & 0.01 & 0.05 \\
         Dimer & 0.002 & 0.1 & 0.2 \\
         Single Atom & 0.0001 & 0.001 & 0.05 \\
         \bottomrule
    \end{tabular}
    \label{tab:si_sigmas}
\end{table}

\section{Crystalline Properties}

\begin{table}[H]
    \caption{Formation energies of the crystalline carbon phases computed with GAP-20 and the tested empirical models compared to those from DFT. In the first table, absolute values of the formation energies relative to the gas phase atom are given, with errors relative to the DFT value in brackets. In the second part of the table, the values for the formation energies are instead given relative to graphite.}
    \begin{adjustbox}{center} 
        \begin{tabular}{lcccccc}
             \toprule
			 & \multicolumn{6}{c}{Formation Energy [eV] (\% Error)} \\
			 \cmidrule(r){2-7}
			 & {DFT} & {GAP-20} & {Tersoff} & {LCBOP} & {REBO-II} & {AIREBO}\\
			 \midrule
			 Graphite           & -8.98 &  -8.97 (0.1) & -7.40 (18) & -7.38 (21) & -7.40 (18) & -7.48 (17)  \\
			 
			 Diamond            & -8.85 & -8.85 (0.1) & -7.37 (18) & -7.35 (18) & -7.37 (18) & -7.46 (17) \\
			 
			 Graphene           & -8.91 & -8.9 (0.1) & -7.40 (18) & -7.35 (18) & -7.40 (18) & -7.43 (17) \\
			 
			 Hexagonal Diamond  & -8.85 & -8.84 (0.1) & -7.37 (17) & -7.34 (18) & -7.37 (17) & -7.30 (18)\\ 
			 
			 Nanotube-(9, 9)    & -8.73 & -8.74 (0.1) & -7.28 (17) & -7.27 (17) & -7.26 (17) & -7.33 (16)  \\
			 
			 Nanotube-(9, 0)    & -8.75 & -8.75 (0.0) & -7.20 (18) & -7.16 (19) & -7.18 (19) & -7.23 (18) \\
			 
			 C60 Fullerene      & -8.53 & -8.51 (0.2) & -6.73 (22) & -6.93 (19) & -6.84 (20) & -6.81 (21) \\
			 
             C100 Fullerene     & -8.52 & -8.60 (0.9) & -6.94 (20) & -7.06 (19) & -7.00 (20) & -7.02 (19)\\
             \midrule
             			 & \multicolumn{6}{c}{Formation Energy Rel. Graphite [eV] (\% Error)} \\
			 \cmidrule(r){2-7}
			 Graphite           & 0	     &	0          &  0     	 &	0          &  0	         & 0   \\
			 
			 Diamond            & 0.12  & 	0.12 (1.2)  &  0.03 (67) & 0.03 (67) & 0.03 (67) & 0.02 (78) \\
			 
			 Graphene           & 0.07	 &  0.07 (0.4)  &  0.00 (100) & 0.03 (100) & 0.00 (100) & 0.05 (29)  \\
			 
			 Lonsdaleite        & 0.13	 &  0.13 (1.6)   &  0.03 (75) & 0.04 (67) & 0.03 (75) & 0.18 (50)  \\ 
			 
			 Nanotube-(9, 9)    & 0.25  &	0.22 (7.5)  &  0.12 (57) & 0.11 (61) & 0.14 (50) & 0.15 (46) \\
			 
			 Nanotube-(9, 0)    & 0.22	 & 0.22	(1.8)    &  0.20 (9) &  0.22 (0)  &  0.22 (0) & 0.25 (14) \\
			 
			 C60 Fullerene      & 0.45  & 0.45	(2.3)	   &  0.67 (49) &  0.45 (0)  & 0.56 (24)  & 0.67 (49)\\
			 
             C100 Fullerene     & 0.35  & 0.37	(5.7)	   &  0.46 (31) &  0.32 (9)  & 0.40 (14) & 0.46 (31) \\
             \bottomrule
        \end{tabular}
    \end{adjustbox}
    \label{tab:crystalline_properties}
\end{table}

\begin{figure}
    \centering
    \includegraphics[width=\textwidth]{Supplemental_Figures/Nanotubes_Formation_Energies_Annotated.png}
    \caption{Formation energies of zig-zag and armchair chirality carbon nanotubes compared to those from DFT. (A) Zig-zag nanotubes formation energies. (B) Armchair nanotubes formation energies. (C) Zig-zag nanotubes formation energies given relative to the formation energy of graphite for each model. (D) Armchair nanotubes formation energies given relative to the formation energy of graphite for each model.}
    \label{fig:supplementary_nanotubes_e_form}
\end{figure}

\newpage

\section{Graphene Phonon Dispersion Curves}

\begin{figure}[H]
    \centering
    \subfigure[]{
    \includegraphics[width=0.35\textwidth]{Supplemental_Figures/Graphene_Phonon/Graphene_Phonons_V9_2b_3b.png}
    }
    \subfigure[]{
    \includegraphics[width=0.35\textwidth]{Supplemental_Figures/Graphene_Phonon/Tersoff.png}
    }
    \subfigure[]{
    \includegraphics[width=0.35\textwidth]{Supplemental_Figures/Graphene_Phonon/LCBOP.png}
    }
    \subfigure[]{
    \includegraphics[width=0.35\textwidth]{Supplemental_Figures/Graphene_Phonon/REBO.png}
    }
    \subfigure[]{
    \includegraphics[width=0.35\textwidth]{Supplemental_Figures/Graphene_Phonon/AIREBO.png}
    }
    \subfigure[]{
    \includegraphics[width=0.35\textwidth]{Supplemental_Figures/Graphene_Phonon/Amorphous_Carbon.png}
    }
    \caption{Phonon dispersion curves for graphene calculated with (A) GAP-20 (B) Tersoff Potential (C) LCBOP (D) REBO (E) AIREBO (F) GAP-17. DFT (optB88-vdW) Reference data are shown in red, while model test data are shown in black. Note that the in this instance, we compare GAP-17 (trained using the local density approximation (LDA) DFT functional) to optB88-vdW reference data, so some disagreement is inevitable. A  comparison of the functional dependence of these data for graphene can be found in \cite{Rowe2018}.}
    \label{fig:graphene_phonons}
\end{figure}

While the phonon dispersion curves for graphene as calculated with a number of empirical models have been reported previously, they are provided here in figure \ref{fig:graphene_phonons} again for completeness - we have additionally computed the phonon dispersion relation for graphene as predicted with the existing GAP-17 potential. Many of the potentials achieve a reasonably good description of the shape of the low frequency phonon modes, however they struggle to describe the spare and energy of higher frequency modes. 

\newpage

\section{Diamond Phonon Dispersion Curves}

\begin{figure}[H]
    \centering
    \subfigure[]{
    \includegraphics[width=0.35\textwidth]{Supplemental_Figures/Diamond_Phonon/Diamond_Phonons_V9_2b_3b.png}
    }
    \subfigure[]{
    \includegraphics[width=0.35\textwidth]{Supplemental_Figures/Diamond_Phonon/Tersoff.png}
    }
    \subfigure[]{
    \includegraphics[width=0.35\textwidth]{Supplemental_Figures/Diamond_Phonon/LCBOP.png}
    }
    \subfigure[]{
    \includegraphics[width=0.35\textwidth]{Supplemental_Figures/Diamond_Phonon/REBO.png}
    }
    \subfigure[]{
    \includegraphics[width=0.35\textwidth]{Supplemental_Figures/Diamond_Phonon/AIREBO.png}
    }
    \subfigure[]{
    \includegraphics[width=0.35\textwidth]{Supplemental_Figures/Diamond_Phonon/Amorphous_Carbon.png}
    }
    \caption{Phonon dispersion curves for diamond calculated with (A) GAP-20 (B) Tersoff Potential (C) LCBOP (D) REBO (E) AIREBO (F) GAP-17. DFT Reference data are shown in red, while model test data are shown in black. Note that the in this instance, we compare GAP-17 (trained using the LDA DFT functional) to optB88-vdW reference data, so some disagreement is inevitable. A  comparison of the functional dependence of the phonon spectrum of graphene can be found in \cite{Rowe2018}.}
    \label{fig:diamond_phonons}
\end{figure}

In figure \ref{fig:diamond_phonons} we provide the calculated phonon modes for diamond using our GAP-20 model, the empirical models tested and the GAP-17 potential. In most cases, the empirical models struggle to predict the shape of the phonon dispersion relations and the phonon band energies are often inaccurate. 

\newpage

\section{Nanotube-(9, 9) Phonon Dispersion Curves}

\begin{figure}[H]
    \centering
    \subfigure[]{
    \includegraphics[width=0.35\textwidth]{Supplemental_Figures/Nanotube_9_9_Phonon/nanotube_9_9_Phonons_V9_2b_3b.png}
    }
    \subfigure[]{
    \includegraphics[width=0.35\textwidth]{Supplemental_Figures/Nanotube_9_9_Phonon/Tersoff.png}
    }
    \subfigure[]{
    \includegraphics[width=0.35\textwidth]{Supplemental_Figures/Nanotube_9_9_Phonon/LCBOP.png}
    }
    \subfigure[]{
    \includegraphics[width=0.35\textwidth]{Supplemental_Figures/Nanotube_9_9_Phonon/REBO.png}
    }
    \subfigure[]{
    \includegraphics[width=0.35\textwidth]{Supplemental_Figures/Nanotube_9_9_Phonon/REBO.png}
    }
    \subfigure[]{
    \includegraphics[width=0.35\textwidth]{Supplemental_Figures/Nanotube_9_9_Phonon/Amorphous_Carbon.png}
    }
    \caption{Phonon dispersion curves and density of states for (9,9) index nanotube calculated with (A) GAP-20 (B) Tersoff Potential (C) LCBOP (D) REBO (E) AIREBO (F) GAP-17. DFT Reference data are shown in red, while model test data are shown in black. Note that the in this instance, we compare GAP-17 (trained using the LDA DFT functional) to optB88-vdW reference data, so some disagreement is inevitable. A  comparison of the functional dependence of the phonon spectrum of graphene can be found in \cite{Rowe2018}.}
    \label{fig:9_9_nanotube_phonons}
\end{figure}

In figure \ref{fig:9_9_nanotube_phonons} we report the phonon dispersion relations for a (9, 9) index carbon nanotube, additionally showing the density of states. It is challenging to make a detailed comparison here regarding specific modes, due to the number and similarity of the various phonon modes. It is more convenient to refer to the density of states in order to draw broad conclusions. Similarly to the case of graphene, the Tersoff potential vastly overestimates the energy of the highest frequency vibrational modes of carbon nanotubes as well as predicting the degeneracy of many phonon modes close to the Brillouin zone edge. The LCBOP, REBO and AIREBO potentials all predict a large spurious peak in the vibrational density of states at high energies. The GAP-17 model correctly predicts the energies of the highest energy modes, but again predicts degenerate vibrational states near the Brillouin zone edge. The GAP-20 model introduced here shows some broadening of the peaks in the vibrational density of states, but generally provides an accurate description of the shape and energies of the phonon dispersion. 

\newpage

\section{Nanotube-(9, 0) Phonon Dispersion Curves}

\begin{figure}[H]
    \centering
    \subfigure[]{
    \includegraphics[width=0.35\textwidth]{Supplemental_Figures/Nanotube_9_0_Phonon/nanotube_9_0_Phonons_V9_2b_3b.png}
    }
    \subfigure[]{
    \includegraphics[width=0.35\textwidth]{Supplemental_Figures/Nanotube_9_0_Phonon/Tersoff.png}
    }
    \subfigure[]{
    \includegraphics[width=0.35\textwidth]{Supplemental_Figures/Nanotube_9_0_Phonon/LCBOP.png}
    }
    \subfigure[]{
    \includegraphics[width=0.35\textwidth]{Supplemental_Figures/Nanotube_9_0_Phonon/REBO.png}
    }
    \subfigure[]{
    \includegraphics[width=0.35\textwidth]{Supplemental_Figures/Nanotube_9_0_Phonon/AIREBO.png}
    }
    \subfigure[]{
    \includegraphics[width=0.35\textwidth]{Supplemental_Figures/Nanotube_9_0_Phonon/Amorphous_Carbon.png}
    }
    \caption{Phonon dispersion curves and density of states for (9,0) index nanotube calculated with (A) GAP-20 (B) Tersoff Potential (C) LCBOP (D) REBO (E) AIREBO (F) GAP-17. DFT Reference data are shown in red, while model test data are shown in black. Note that the in this instance, we compare GAP-17 (trained using the LDA DFT functional) to optB88-vdW reference data, so some disagreement is inevitable. A  comparison of the functional dependence of the phonon spectrum of graphene can be found in \cite{Rowe2018}.}
    \label{fig:9_0_nanotube_phonons}
\end{figure}

In figure \ref{fig:9_0_nanotube_phonons} we present the phonon dispersion curves and vibrational density of states for a (9, 0) index carbon nanotube as predicted by GAP-20, a number of empirical potentials and the GAP-17 model. As in the case of the (9, 9) index carbon nanotube, the Tersoff potential overestimates the energy of the high energy phonon modes by approx. 100 meV, while the REBO, LCBOP and AIREBO potentials have a spurious peak in the vibrational density of states at high energies. In this instance, the GAP-17 model underestimates the energy of the phonon modes, but predicts the correct shape for the density of states. The GAP-20 model shows some broadening of the peaks in the vibrational density of states, but is generally accurate for both energies and dispersion shape. 

\newpage

\section{Graphene bilayer separation energy}

\begin{figure}[H]
    \centering
    \includegraphics[width=0.7\textwidth]{Supplemental_Figures/Graphite_Combined_Pull.png}
    \caption{Energy required to move a bound system of two graphene sheets out to large separation.}
    \label{fig:graphene_layer_sep}
\end{figure}

Figure \ref{fig:graphene_layer_sep} shows the predicted energy as the distance between two graphene sheets is changed. It provides a useful test of how well a potential might capture the long-ranged Van der Waals interactions, which are important for modelling layered materials like graphite. GAP-20 accruately reproduces the shape and depth of the binding curve out to long separation as do both the AIREBO and LCBOP potentials, although the latter does underbind the two sheets. GAP-17 strongly overbinds the two sheets at very short distances. Due to their extremely short cutoffs, the REBO and Tersoff potentials predict no interaction between the graphene sheets even at very short distances. 

\section{Force and Energy Errors of the Potential}

\begin{figure}[H]
    \centering
    \includegraphics[width=\textwidth]{Supplemental_Figures/Force_Errors_1.png}
    \caption{Force Errors (1/3) (Left to right) Correlation between GAP-20 and DFT (optB88-vdW) forces, errors in force prediction for GAP-20, correlation between GAP-20 and DFT (optB88-vdW) energies and errors in energy prediction for GAP-20. (Top to bottom) data for bulk amorphous, dimer, SACADA \cite{Hoffmann2016} database, fullerenes, nanotubes and bulk crystalline structues.}
    \label{fig:force_errors_1}
\end{figure}

\begin{figure}[H]
    \centering
    \includegraphics[width=\textwidth]{Supplemental_Figures/Force_Errors_2.png}
    \caption{Force Errors (2/3) (Left to right) Correlation between GAP-20 and DFT (optB88-vdW) forces, errors in force prediction for GAP-20, correlation between GAP-20 and DFT (optB88-vdW) energies and errors in energy prediction for GAP-20. (Top to bottom) data for defective carbon, graphite, carbon surfaces, crystalline RSS data \cite{Deringer2017a}, liquid carbon and liquid-solid interfaces.}
    \label{fig:force_errors_2}
\end{figure}

\begin{figure}[H]
    \centering
    \includegraphics[width=\textwidth]{Supplemental_Figures/Force_Errors_3.png}
    \caption{Force Errors (3/3) (Left to right) Correlation between GAP-20 and DFT (optB88-vdW) forces, errors in force prediction for GAP-20, correlation between GAP-20 and DFT (optB88-vdW) energies and errors in energy prediction for GAP-20. (Top to bottom) data for graphene, diamond, the isolated gas-phase atom, bilayer graphene and amorphous carbon surfaces.}
    \label{fig:force_errors_3}
\end{figure}

\section{Optimisation of SOAP Descriptor}

\begin{figure}[h]
    \centering
    \includegraphics[width=0.7\textwidth]{Supplemental_Figures/SOAP_Force_Sigma.png}
    \caption{Convergence of mean absolute force errors on the test set for an l=8, n=8 SOAP descriptor with a 4.2 \AA cutoff and 6000 sparse points with respect to the $\sigma_{\mathrm{force}}$ parameter used in GAP training.}
    \label{fig:soap_force_sigma}
\end{figure}

Figure \ref{fig:soap_force_sigma} shows the convergence of the force error as a function of the selected value for $\sigma_{\mathrm{force}}$. The accuracy of the potential is strongly affected if values larger than approx 0.01 eV \AA$^{-1}$ are used, and begin to increase again for very small values as over-fitting becomes problematic. In general, one wishes to use the largest value possible without negatively impacting the quality of the potential, thereby minimising the effects of over-fitting and maximising the resistance of the potential to noise or other possible artefacts in the training data. 

\section{Random Structure Search}

In addition to showing the error of GAP-20 compared to DFT when computing the energies of the GAP-RSS structures, we here also plot in sketch-map representation (fig \ref{fig:supp_rss} the energies of the structures themselves. 

\begin{figure}[h]
    \centering
    \includegraphics[width=\textwidth]{Supplemental_Figures/RSS_Energy_Annotated.png}
    \caption{Sketch-map representation of GAP-RSS structures, the energy of each structure is indicated by the colour of the point. The density of the population of structures in the GAP-20 training data are indicated by the black contour lines.}
    \label{fig:supp_rss}
\end{figure}

\section{Cost of the Potential}
The primary benefit to using a Gaussian approximation potential over direct \textit{ab initio} simulation is the significant reduction in the cost of simulations.
We have previously commented on the relative cost of direct \textit{ab initio} simulation compared to GAP models, density functional tight binding and a number of empirical models for a small 200 atom graphene system \cite{Rowe2018}.
Here, we compare the cost of our GAP model to that of direct \textit{ab initio} simulation with VASP and to the LCBOP empirical many body potential for carbon. 
Although there are many empirical potentials for carbon available, for our illustrative purposes here they perform effectively the same; the LCBOP potential could be substituted for any number of other empirical potentials without affecting our conclusions here.
Simulations were performed in the NVT ensemble at $1000 \, \mathrm{K}$ for diamond lattices ranging from 8 to 5832 atoms.
These simulations were performed using 72 cores spread over 3 nodes on the Thomas cluster, the UK National Tier 2 High Performance Computing Hub for Materials and Molecular Modelling.
In the case of GAP and LCBOP simulations, trajectories of 20,000 steps were generated, for VASP \textit{ab initio} molecular dynamics simulations, trajectories of 10 steps were generated.
\textit{Ab initio} molecular dynamics simulations were performed at the gamma point, the electronic convergence criterion was set to $10^{-4} \, \mathrm{eV}$, all other settings were fixed as described for the generation of the training data.
We note that for very small system sizes, the empirical models will spend a considerable portion of their time on communication rather than on calculation of the pair potential, however this is unavoidable if we wish to perform comparable simulations on larger cells with DFT. 

\begin{figure}[H]
    \centering
    \includegraphics[width=0.5\textwidth]{./Figures/V7_2_3000/Timings.png}
    \caption{Efficiency of the GAP model for simulations of various sizes. Note the logarithmic scale of the y-axis.}
    \label{fig:timings}
\end{figure}

As can be seen from figure \ref{fig:timings}, the GAP model is many orders of magnitude cheaper than \textit{ab initio} molecular dynamics, even for relatively modest system sizes. 
Furthermore, since the scaling of the GAP model follows the approximately $\mathrm{N} \mathrm{log}(\mathrm{N})$ scaling of empirical MD, while that of DFT scales with the cube of the number of electrons, the advantage to using the GAP model increases dramatically for larger systems.
In fact, simulations using DFT for systems larger than 2744 atoms was not possible with the chosen computational setup. 
While GAP model does, have a much larger prefactor to this scaling than any of the empirical models tested, it remains entirely feasible to use it to simulate systems of tens of thousands of atoms for nanoseconds at a time.

\bibliographystyle{ieeetr}
\bibliography{Patrick_Rowe_Bibliography}